\documentstyle[11pt,appb,epsfig,axodraw]{article} 
\def\refname{\large{\bf References}}

\newcommand{\ks}{k\hspace{-0.52em}/\hspace{0.1em}}

\newcommand{\qs}{q\hspace{-0.5em}/\hspace{0.1em}}

\begin{document}
\title{CHARM PRODUCTION IN DEEP INELASTIC\\ 
       LEPTON-HADRON SCATTERING\thanks{lectures presented at the 
          XXXVIIth Cracow School of Theoretical Physics, 30th May - 
          10th June 1997, Zakopane, Poland.}
% you can use \\ to break lines
}
\author{W. L. van Neerven
\address{Instituut--Lorentz, University of Leiden, POB 9506, 2300 RA Leiden
, The Netherlands}
}
\maketitle
\begin{abstract}
A review is given of the QCD corrections to charm quark production in
deep inelastic electron-proton scattering. An outline of the
computation of the virtual
photon-parton subprocesses, from which one obtains the heavy quark 
coefficient functions, is given. The dominant production mechanisms
are discussed. Further we show that the asymptotic heavy quark coefficient
functions, computed in the limit $Q^2 \gg m^2$,
can be derived using the operator
product expansion technique. Further we present the various schemes proposed
in the literature to describe the charm component of the structure function
and compare them with the most recent data from the experiments carried
out at HERA.
\end{abstract}
\section{Introduction}
The study of heavy quarks and their decay and production mechanisms provides
us with important insights in the standard model of the electroweak and
strong interactions. In this model the heavy flavours are given by the charm
(c), bottom (b) and the top (t) quarks. Because of the confinement property
in the theory of the strong interactions, given by quantum chromodynamics
(QCD), these quarks can only be observed via the heavy baryons and mesons
in which they are confined. Since mesons are more copiously produced than
baryons we concentrate on the former particles. Here one can 
distinguish between the following type of mesons.
\begin{itemize}
\item[1] open heavy quark mesons\\[3mm]
In this case the heavy quark is accompanied by a lighter quark mostly
represented by the up (u), down (d) and strange (s) quarks. Examples are:
\begin{itemize}
\item[a] open charm e.g. $D_u=\bar u c$, $D_d=\bar d c$, $D_s=\bar s c$.
\item[b] open bottom e.g.
$B_u=\bar u b$, $B_d=\bar d b$, $B_s=\bar s b$, $B_c= \bar c b$
\end{itemize}
including their anti-mesons.
\item[2] hidden heavy quark mesons\\[3mm]
Here the heavy quark is always bound to its anti-quark. Examples are:
\begin{itemize}
\item[a] hidden charm e.g. ${\cal J}/\psi = \bar c c$
\item[b] hidden bottom e.g. ${\mit\Upsilon} = \bar b b$
\end{itemize}
including the higher excitations.
\end{itemize}
In these lectures we limit ourselves to open heavy quark production.
The mesons are observed via their electroweak decays. 
An example is the decay of
the meson $D_u$ given by $D_u \rightarrow K^{-}+\mu^{+}+\nu_{\mu}$ which
proceeds via the partonic reaction $c \rightarrow s + \mu^{+}+\nu_{\mu}$
where $\mu^{+},\nu_{\mu}$ emerge from the virtual $W^{+}$-boson which is
exchanged between the quarks and the lepton pair.\\
There are numerous reasons why open heavy quark mesons are of interest. Here
we will mention some of them (for a review see \cite{ali}).
\begin{itemize}
\item[a] The observation of rare decays in particular those of the B-mesons.
With rare we mean decays that are suppressed in the standard model.
\item[b] The measurement of the Cabibbo-Kobayashi-Maskawa matrix 
elements denoted
by $V_{ij}$ ($i,j=u,d \cdots t$) via the production or decay of the heavy 
quark mesons. A classical example is the measurement of the quantity $V_{cs}=
\cos \theta_c$ in the process $e + p \rightarrow \nu_e + D + 'X'$ ($'X'$ is any 
inclusive final state). For this reaction the dominant partonic subprocess
is given by $e + s \rightarrow \nu_e + c$ which proceeds via the exchange of
a W-boson.
\item[c] The study of $D\bar D$ and $B\bar B$-mixing in connection with CP-
violation in the B-system.
\item[d]
The study of production mechanisms of heavy quarks provides us with new tests
of QCD. It also enables us to measure some of the parton densities (see below)
in kinematical regions which are not accessible in other types of processes.
An example is the gluon density which can be measured via charm production 
at the HERA-collider.
\end{itemize}
In these lectures we will only study those production mechanisms of heavy
quarks where the methods of perturbative QCD can be used. Reactions where
only nonperturbative methods can be applied like e.g.,
diffraction will not be treated here.\\

In perturbative QCD physical quantities can be expanded in the strong coupling
constant $\alpha_s(\mu^2)$, where the scale $\mu$ has to be large, provided
we are dealing with so called hard processes. The latter are described by
the parton model including higher order QCD corrections. The hard
processes are characterized by the property that all kinematical invariants, 
on which the quantities depend, become asymptotic whereas their mutual ratios
are fixed. These ratios should neither become zero nor infinite. In the case
of heavy quark production the above condition implies that the heavy quark mass
$m$ should become asymptotic too. Unfortunately these types of processes 
cannot be completely described by perturbative methods. All cross sections also
contain nonperturbative parts. The latter are represented by the parton 
densities and the fragmentation functions whose properties will be discussed 
below.\\
Let us first enumerate the various reactions, including the basic partonic 
subprocesses, in which the heavy mesons are produced in a semi inclusive way.
\begin{itemize}
\item[1] hadron-hadron collisions\\[3mm]
Example: $ P + \bar P \rightarrow B + \bar B + 'X'$\\[3mm]
Lowest order partonic subprocesses: $g + g \rightarrow b + \bar b$, 
$q + \bar q  \rightarrow b + \bar b$ 
\item[2] lepton-hadron collisions\\[3mm]
Example: $e + P \rightarrow D + \bar D + 'X'$\\[3mm]
Lowest order partonic subprocess: $\gamma^* + g \rightarrow c + \bar c$ 
\item[3] photon-hadron collisions\\[3mm]
Example: $\gamma + P \rightarrow D + \bar D + 'X'$\\[3mm]
Lowest order partonic subprocess: $\gamma + g \rightarrow c + \bar c$ 
\item[4] photon-photon collisions\\[3mm]
Example: $\gamma + \gamma \rightarrow D + \bar D + 'X'$\\[3mm]
Lowest order partonic subprocess: $\gamma + \gamma \rightarrow c + \bar c$ 
\item[5] electron-positron collisions\\[3mm]
Example: $e^{+} + e^{-} \rightarrow B + \bar B + 'X'$\\[3mm]
Lowest order partonic subprocess: $e^{+} + e^{-} \rightarrow b + \bar b$
\end{itemize} 
%---------------
% fig1
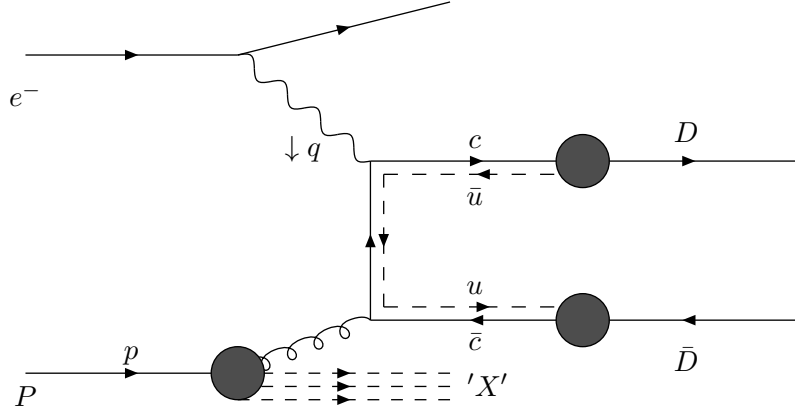
\begin{figure}
\begin{center}
  \begin{picture}(290,160)(0,0)
    \ArrowLine(0,20)(80,20)
    \Gluon(80,20)(130,40){3}{4}
    \DashArrowLine(80,20)(160,20){5}
    \DashArrowLine(80,15)(160,15){5}
    \DashArrowLine(80,10)(160,10){5}
    \GCirc(80,20){10}{0.3}
    \ArrowLine(210,40)(130,40) 
    \ArrowLine(130,40)(130,100) 
    \ArrowLine(130,100)(210,100) 
    \DashArrowLine(210,95)(135,95){5}      
    \DashArrowLine(135,95)(135,45){5}      
    \DashArrowLine(135,45)(210,45){5}      
    \ArrowLine(290,40)(210,40)
    \ArrowLine(210,100)(290,100) 
    \GCirc(210,40){10}{0.3}
    \GCirc(210,100){10}{0.3} 
    \Photon(80,140)(130,100){3}{4}
    \ArrowLine(0,140)(80,140)
    \ArrowLine(80,140)(160,160)
    \Text(105,110)[t]{$\downarrow q$}
    \Text(0,130)[t]{$e^-$}
    \Text(40,30)[t]{$p$}
    \Text(0,15)[t]{$P$}
    \Text(175,20)[t]{$'X'$}
    \Text(250,115)[t]{$D$}
    \Text(250,30)[t]{$\bar D$}
    \Text(170,35)[t]{$\bar c$}
    \Text(170,110)[t]{$c$}
    \Text(170,90)[t]{$\bar u$}
    \Text(170,55)[t]{$u$}
  \end{picture} 
  \caption[]{Production of the mesons $D$ and $\bar D$ in deep inelastic 
             electron-proton scattering via the photon-gluon fusion
             mechanism.}
  \label{fig:1}
\end{center}
\end{figure}
%---------------------------
Notice that in reaction $4$ one or both photons can be virtual (indicated by
$*$). In these lectures we only concentrate on reaction $2$
where the charm\\
(anti-) quark is produced in deep inelastic lepton-hadron
scattering. 
We show this process in more detail in Fig. \ref{fig:1}, 
where the momenta of the virtual photon and the proton are 
denoted by $q$ and $p$ respectively. 
Notice that $q$ is spacelike (i.e. $q^2 =-Q^2 < 0$). Further we 
assume that $Q^2$ is small enough so that the neutral current
process with the exchange of
a Z-boson is suppressed. The process becomes deep inelastic when $Q^2$ and
$p\cdot q$ are large whereas the scaling variable $x= Q^2/2p\cdot q$ 
is kept fixed
(i.e. $0<x<1$ but $x\not =0$ and $x \not =1$). When these kinematical
conditions are satisfied then according to the parton model the proton
can be viewed as a collection of free quarks and gluons (called partons).
Each of them can be involved in the hard scattering without being influenced
by the other partons which are called spectators. The latter produce the
hadronic final state $'X'$. Integration over the momenta
and summation over the quantum numbers of the spectators provides us with
the probability that the gluon in Fig. \ref{fig:1}
emerges from the proton with the
fraction $z$ of the proton's momentum. This probability is given by the
parton density indicated by $f_g^P(z)$. This function can only be computed
using nonperturbative methods in QCD. Unfortunately these methods are not
available yet so that one either has to resort to models for
$f_g^P(z)$ or one has to obtain this density 
from the data. The same holds for the quark parton densities denoted by
$f_q^P(z)$. The incoming gluon in Fig. \ref{fig:1} is involved in a hard 
scattering
with the photon so that a charm anti-charm quark pair is produced 
in the final state. The charm and anti-charm quark
pick up from the vacuum a light anti-quark and quark 
respectively. The fusion of the charm quark with the light anti-quark leads
to the production of a D-meson. This process is described by the fragmentation
function ${\cal D}_c^D(z)$ where $z$ 
denotes the fraction of the momentum of the
charm quark carried away by the D-meson. An analogous description holds for
the production of the anti D-meson coming from the anti-charm and the light
quark. The light (anti-) quarks, which are called spectators, have very low
momenta so that they cannot change the magnitude as well as the direction of
the three-momentum of the charm quark. Therefore the momentum of the charm
quark can be reconstructed from the momentum of the D-meson. Like the parton
densities the fragmentation functions can be only determined by using
nonperturbative methods. However the latter are not at that stage that one
can compute these functions. Therefore they can either be determined by models
or they have to be extracted from the data. In spite of the poor knowledge 
about these
functions one can assume that the parton densities $f_k^P(z)$ ($k=q,g$) and
the fragmentation functions ${\cal D}_k^a(z)$ ($a=D,B; k=c,b$) are universal
and process independent. The former only depend on the type of parton (q or g)
and the type of hadron from which the parton emerges. The same holds for
the fragmentation function which only depends on the heavy quark and the type
of meson into which the former fragments. This universality remains unaltered
even after QCD corrections to the hard processes have been included. It
means that when these phenomenological functions are obtained from a certain 
process $A$ one can use them as input for process $B$ in order to make
absolute predictions for the cross section of the latter reaction. This
statement is only correct if the QCD corrections are carried out up to the
same order in both processes.
%----------------------
%fig2
\begin{figure}
\begin{center}
  \begin{picture}(170,130)(0,0)
    \ArrowLine(0,20)(80,30)
    \ArrowLine(80,40)(160,80)
    \DashArrowLine(80,30)(160,35){5}
    \DashArrowLine(80,25)(160,25){5}
    \DashArrowLine(80,20)(160,15){5}
    \GCirc(80,30){20}{0.3}
    \Photon(50,100)(80,40){3}{7}
    \ArrowLine(0,100)(50,100)
    \ArrowLine(50,100)(100,120)
    \Text(0,110)[t]{$e^-$}
    \Text(100,130)[t]{$e^-$}
    \Text(0,15)[t]{$P$}
    \Text(160,75)[t]{$c~(\bar c)$}
    \Text(60,70)[t]{$V$}
    \Text(175,30)[t]{$'X'$}
    \Text(40,20)[t]{$p$}
    \Text(25,115)[t]{$l_1$}
    \Text(75,125)[t]{$l_2$}
    \Text(80,70)[t]{$\downarrow q$}
    \Text(120,55)[t]{$p_1$}
  \end{picture}
  \caption[]{Kinematics of charm production in deep inelastic
             electron-proton scattering} 
  \label{fig:2}
\end{center}
\end{figure}
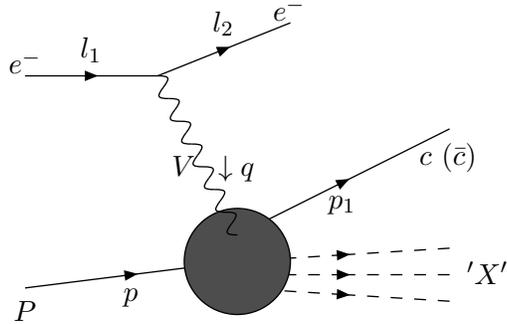
%----------------------------
\section{Electroproduction of heavy quarks}
\begin{figure}
\begin{center}
  \begin{picture}(130,65)
  \Gluon(0,0)(30,15){3}{7}
  \ArrowLine(60,15)(30,15)
  \ArrowLine(30,15)(30,45)
  \ArrowLine(30,45)(60,45)
  \Photon(0,60)(30,45){3}{7}
    \Text(9,0)[t]{$g$}
    \Text(9,65)[t]{$\gamma$}
    \Text(65,48)[t]{$c$}
    \Text(65,18)[t]{$\bar c$}
  \end{picture}
\hspace*{1cm}
  \begin{picture}(130,65)
  \Gluon(0,0)(30,15){3}{7}
  \ArrowLine(30,15)(60,15)
  \ArrowLine(30,45)(30,15)
  \ArrowLine(60,45)(30,45)
  \Photon(0,60)(30,45){3}{7}
    \Text(9,0)[t]{$g$}
    \Text(9,65)[t]{$\gamma$}
    \Text(65,48)[t]{$\bar c$}
    \Text(65,18)[t]{$c$}
  \end{picture}
\vspace*{5mm}
  \caption{Feynman diagrams for the lowest-order photon-gluon
           fusion process contributing to the coefficient functions
           $H_{i,g}^{(1)}$.} 
  \label{fig:3}
\end{center}
\end{figure}
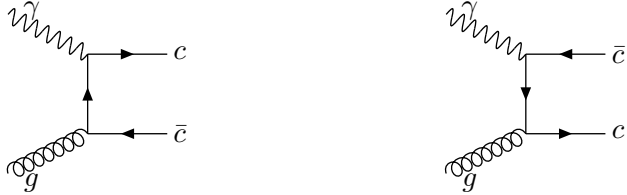

Charm quark production in deep inelastic electron-proton scattering proceeds
via the following reaction (see Fig. \ref{fig:2})
\begin{eqnarray}
\label{eqn:1}
e^-(l_1) + P(p) \rightarrow e^-(l_2) + c (p_1)\,( \bar c(p_1) ) +'X'\,.
\end{eqnarray}
Here $'X'$ denotes any inclusive hadronic state which means that we have
summed over
all quantum numbers and integrated over all momenta of the hadrons belonging
to this state. Further we consider neutral current processes only so that
the intermediate vector boson $V$ either stands for the photon or for the 
Z-boson. The momentum transfer in the above process is spacelike so that we
have
\begin{eqnarray}
\label{eqn:2}
q^2 \equiv - Q^2 < 0 \qquad   \mbox{with}  \qquad q=l_1-l_2\,.
\end{eqnarray}
In the subsequent part of these lectures we will assume that $Q^2 \ll M_Z^2$
so that the above reaction is dominated by the one-photon exchange mechanism.
The computation of the cross section of reaction (\ref{eqn:1}) involves a 
five-dimensional integral. The kinematical integration variables are 
\cite{lrsn1}
\begin{eqnarray}
\label{eqn:3}
S=(p+q)^2 \quad , \quad
\cos \Phi = \frac{(\vec l_1 \times \vec l_2).(\vec p \times \vec p_1)}
{\mid \vec l_1 \times \vec l_2 \mid . \mid \vec p \times \vec p_1)\mid}\,,
\end{eqnarray}
\begin{eqnarray}
\label{eqn:4}
T_1=T-m^2=(p-p_1)^2-m^2 \quad , \quad  U_1=U-m^2=(q-p_1)^2-m^2 \,.
\end{eqnarray}
Further we define the scaling variables
\begin{eqnarray}
\label{eqn:5}
x=\frac{Q^2}{2\,p\cdot q} \quad , \quad y=\frac{p\cdot q}{p\cdot l_1}\,.
\end{eqnarray}
Integration over $\Phi$ yields the following cross section
\begin{eqnarray}
\label{eqn:6}
&&\frac{d^4 \sigma (T_1,U_1,x,y)}{dx\,dy\,dT_1\,dU_1} =
\nonumber\\
&& \frac{\alpha}{2\pi\,x\,y}
\Big [ 2(1-y) \frac{d^2 \sigma_L(T_1,U_1,x)}{dT_1\,dU_1} +
\{ 1 + (1-y)^2\} \frac{d^2\sigma_T(T_1,U_1,x)}{dT_1\,dU_1} \Big ]\,,
\end{eqnarray}
where $\sigma_L$ and $\sigma_T$ stand for the longitudinal and transverse
photon cross sections respectively. 
In addition to the variables indicated above
$\sigma$ and $\sigma_k$ ($k=T,L$) also depend on $Q^2$ Eq. (\ref{eqn:2}) and the
heavy quark mass $m$. The factor in front of $d^2 \sigma_T$ in 
Eq. (\ref{eqn:6}) reflects the
vector nature of the photon. From $d^2 \sigma_k/dT_1 dU_1$ in Eq. (\ref{eqn:6})
one can compute the double differential cross sections \cite{lrsn2}
\begin{eqnarray}
\label{eqn:7}
\frac{d^2\sigma_k}{dY\,d\mid \vec p_{1\bot}\mid}\quad , \quad
 \frac{d^2\sigma_k}{dX_f\,d\mid \vec p_{1\bot}\mid}\,.
\end{eqnarray}
The rapidity $Y$ and the momentum fraction $X_f$ are defined by
\begin{eqnarray}
\label{eqn:8}
Y=\frac{1}{2} \ln \Big (\frac{E_1+\mid\vec p_{1l}\mid}{E_1-\mid \vec p_{1l}\mid}
\Big) \quad , \quad X_f= \frac{\mid \vec p_{1l}\mid}{\mid \vec p_l\mid }\,.
\end{eqnarray}
In the above equations the longitudinal and the transverse momenta are 
indicated by the indices $l$ and $\bot$ respectively. In the subsequent part
of these lectures we are only interested in the integrated cross sections
given by
\begin{eqnarray}
\label{eqn:9}
\sigma_k(x,Q^2,m^2)=\int dT_1 \int dU_1 \frac{d^2\sigma_k(T_1,U_1,x,Q^2,m^2)}
{dT_1\,dU_1}\,.
\end{eqnarray}
The total longitudinal and transverse cross sections are related to the
corresponding deep inelastic structure functions $F_k$ ($k=T,L$) as follows
\begin{eqnarray}
\label{eqn:10}
\sigma_L(x,Q^2,m^2)&=&\frac{4 \pi^2 \alpha}{Q^2} F_L(x,Q^2,m^2) \, \qquad
\nonumber\\
\sigma_T(x,Q^2,m^2)&=&\frac{4 \pi^2 \alpha}{p\cdot q} F_1(x,Q^2,m^2)\,.
\end{eqnarray}
Instead of $F_1$ the experimentalists measure the structure function 
\begin{eqnarray}
\label{eqn:11}
F_2(x,Q^2,m^2)=2\,x\,F_1(x,Q^2,m^2)+F_L(x,Q^2,m^2)\,.
\end{eqnarray}
With the above definitions the deep inelastic scattering cross section can be
written as
\begin{eqnarray}
\label{eqn:12}
\frac {d^2\sigma}{dx\,dy}=\frac{2 \pi \alpha^2}{Q^2} S_{eP}
 \Big[ \{ 1 + (1-y)^2\}
F_2(x,Q^2,m^2)- y^2 F_L(x,Q^2,m^2) \Big]\,.
\end{eqnarray}
Here $\sqrt S_{eP}$ denotes the centre of mass energy of incoming 
electron-proton 
system. When the charm component of the structure function is under study
we will add an index $c$ to the structure functions so $F_k$ is replaced 
by $F_{k,c}$. Before we continue it is important to emphasize
that perturbative QCD can only predict the $Q^2$-evolution of the structure 
functions but not their $x$-dependence. The latter is partially determined by 
the parton densities which, as we have mentioned before, are of a 
nonperturbative origin.\\
For charm production in perturbative QCD one can distinguish between two 
different production mechanisms. Let us denote the proton state in the Fock
space by
\begin{eqnarray}
\label{eqn:13}
\mid P \rangle &=& a_1 \mid uud \rangle + a_2 \mid u \bar u, uud \rangle
+ a_3 \mid d \bar d, uud \rangle + a_4 \mid s \bar s,uud \rangle 
\nonumber\\
&& + \sum_{H=c}^t a_H \mid H \bar H,uud \rangle\,.
\end{eqnarray}
The two possible production mechanisms are:
\begin{itemize}
\item[A] \underline{Intrinsic heavy quark production}\cite{bhmt}\\[3mm]
Here we have $a_H \not = 0$ for at least one
$H$ with $H=c,b,t$. In the case of intrinsic charm we have $a_c \not =0$
and $a_b =0$, $a_t=0$. The main production mechanism is given by the flavour
excitation process
\begin{eqnarray}
\label{eqn:14}
         \gamma^* + c \rightarrow c \,,
\end{eqnarray}
which is a zeroth order process in the strong coupling
constant  $\alpha_s$.
\item[B] \underline{ Extrinsic heavy quark production}\cite{wit}\\[3mm]
Here we have $a_H=0$ for $H=c,b,t$. The dominant production mechanism is now
given by the photon-gluon fusion process. On the Born level (first order in
$\alpha_s$) it is given by (see Fig. \ref{fig:3})
\begin{eqnarray}
\label{eqn:15}
            \gamma^* + g \rightarrow c + \bar c\,.
\end{eqnarray}  
\end{itemize}
Notice that in the case of extrinsic charm production only light quarks (u,d,s)
and the gluon can appear in the initial state of the partonic processes since
the probability that a charm quark emerges from the proton is zero. In these 
lectures we will limit ourselves to extrinsic charm production because the
recent experiments at HERA \cite{h1},\cite{zeus} indicate that the data favour 
this production mechanism so that $f_c^P=0$.
%--------------------------------
%fig3
%-------------------------------- 
\section{Exact heavy quark coefficient functions up to
order $\alpha_s^2$}
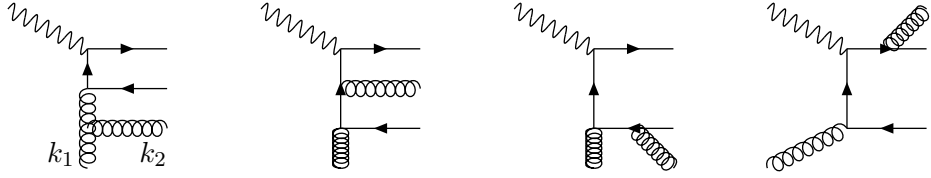
\begin{figure}
\begin{center}
  \begin{picture}(60,60)
  \Gluon(30,0)(30,30){3}{7}
  \ArrowLine(60,30)(30,30)
  \ArrowLine(30,30)(30,45)
  \ArrowLine(30,45)(60,45)
  \Gluon(30,15)(60,15){3}{7}
  \Photon(0,60)(30,45){3}{7}
    \Text(20,10)[t]{$k_1$}
    \Text(55,10)[t]{$k_2$}
  \end{picture}
\hspace*{1cm}
%  \hfill
  \begin{picture}(60,60)
  \Gluon(30,0)(30,15){3}{7}
  \ArrowLine(60,15)(30,15)
  \ArrowLine(30,15)(30,45)
  \ArrowLine(30,45)(60,45)
  \Gluon(30,30)(60,30){3}{7}
  \Photon(0,60)(30,45){3}{7}
\end{picture}
\hspace*{1cm}
%  \hfill
  \begin{picture}(60,60)
  \Gluon(30,0)(30,15){3}{7}
  \ArrowLine(60,15)(30,15)
  \ArrowLine(30,15)(30,45)
  \ArrowLine(30,45)(60,45)
  \Gluon(45,15)(60,0){3}{7}
  \Photon(0,60)(30,45){3}{7}
\end{picture}
\hspace*{1cm}
%  \hfill
  \begin{picture}(60,60)
  \Gluon(0,0)(30,15){3}{7}
  \ArrowLine(60,15)(30,15)
  \ArrowLine(30,15)(30,45)
  \ArrowLine(30,45)(60,45)
  \Gluon(45,45)(60,60){3}{7}
  \Photon(0,60)(30,45){3}{7}
\end{picture}
\vspace*{5mm}
  \caption{The  bremsstrahlungsprocess $\gamma^* + g \rightarrow
           c + \bar c + g$ contributing to the coefficient functions
           $H_{i,g}^{(2)}$.}
  \label{fig:5}
\end{center}
\end{figure}

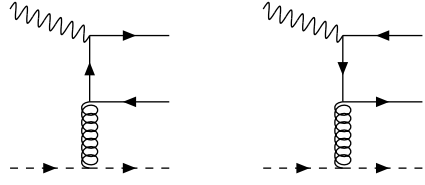
\begin{figure}
\begin{center}
  \begin{picture}(60,60)
  \DashArrowLine(0,0)(30,0){3}
  \DashArrowLine(30,0)(60,0){3}
  \Gluon(30,0)(30,25){3}{7}
  \ArrowLine(60,25)(30,25)
  \ArrowLine(30,25)(30,50)
  \ArrowLine(30,50)(60,50)
  \Photon(0,60)(30,50){3}{7}
  \end{picture}
\hspace*{1cm}
%  \hfill
  \begin{picture}(60,60)
  \DashArrowLine(0,0)(30,0){3}
  \DashArrowLine(30,0)(60,0){3}
  \Gluon(30,0)(30,25){3}{7}
  \ArrowLine(30,25)(60,25)
  \ArrowLine(30,50)(30,25)
  \ArrowLine(60,50)(30,50)
  \Photon(0,60)(30,50){3}{7}
\end{picture}
\vspace*{5mm}
  \caption{The Bethe-Heitler process $\gamma^* + q \rightarrow
           c + \bar c + q$ contributing to the coefficient functions
           $H_{i,q}^{(2)}$.}
  \label{fig:6}
\end{center}
\end{figure}

The calculation of the partonic cross sections denoted by $\hat \sigma_{i,k}$
is straightforward (see \cite{wit}). After using the 
same relations as presented
for the structure functions in Eq. (12) we obtain the heavy quark coefficient
functions represented by $H_{i,k}$ ($i=2,L;k=q,g$). The lowest order
contribution to the structure functions corresponding to the graphs in 
Fig. (\ref{fig:3}) can be written as
\begin{eqnarray}
\label{eqn:16}
F_{i,c}^{(1)}(x,Q^2,m^2)&=&e_c^2 \int_x^{z_{\rm max}} \frac{dz}{z}
 \hat f_g^{\rm S}(x/z)
\hat H_{i,g}^{(1)}(z,Q^2,\hat m^2,\hat \alpha_s) 
\nonumber\\
&\equiv& e_c^2 \, \hat f_g^{\rm S} \otimes \hat H_{i,g}^{(1)} \,,
\end{eqnarray}
with $z_{\rm max}=Q^2/(Q^2+4m^2)$ 
and $e_c$ denotes the charge of the heavy quark 
(here $e_c=2/3$). 
The quantities $\hat m$, $\hat \alpha_s$ and $\hat f_g$ stand for the 
bare mass, bare coupling
constant and bare gluon density respectively. The meaning of the latter 
function will be discussed later. Further the order $\alpha_s^n$ contribution 
to the
heavy quark coefficient function will be denoted by $H_{i,k}^{(n)}$.
Finally the superscript on the gluon density indicates that we are dealing
with a singlet (S) quantity with respect to the flavour symmetry 
transformations. Here the flavour symmetry group is given by SU$(n_f)$. In the
case of charm production we choose $n_f=3$.\\
%---------------------------
%fig4
\begin{figure}
\begin{center}
  \begin{picture}(60,60)
  \Gluon(0,0)(30,15){3}{7}
  \ArrowLine(60,15)(30,15)
  \ArrowLine(30,15)(30,45)
  \ArrowLine(30,45)(60,45)
  \Gluon(45,45)(45,15){3}{7}
  \Photon(0,60)(30,45){3}{7}
  \end{picture}
\hspace*{1cm}
%  \hfill
  \begin{picture}(60,60)
  \Gluon(0,0)(30,15){3}{7}
  \ArrowLine(60,15)(30,15)
  \ArrowLine(30,15)(30,45)
  \ArrowLine(30,45)(60,45)
  \Gluon(30,30)(45,15){3}{7}
  \Photon(0,60)(30,45){3}{7}
\end{picture}
\hspace*{1cm}
%  \hfill
  \begin{picture}(60,60)
  \Gluon(0,0)(30,15){3}{7}
  \ArrowLine(60,15)(30,15)
  \ArrowLine(30,15)(30,45)
  \ArrowLine(30,45)(60,45)
  \Gluon(30,30)(45,45){3}{7}
  \Photon(0,60)(30,45){3}{7}
\end{picture}
\hspace*{1cm}
%  \hfill
  \begin{picture}(60,60)
  \Gluon(0,0)(30,15){3}{7}
  \ArrowLine(60,15)(30,15)
  \ArrowLine(30,15)(30,45)
  \ArrowLine(30,45)(60,45)
  \Gluon(15,7)(45,45){3}{7}
  \Photon(0,60)(30,45){3}{7}
    \Text(9,0)[t]{$k_1$}
    \Text(45,35)[t]{$k$}
\end{picture}
\vspace*{5mm}
  \caption{Virtual gluon corrections to the proces $\gamma^* + g \rightarrow
           c + \bar c$ contributing to the coefficient functions
           $H_{i,g}^{(2)}$.}
  \label{fig:4}
\end{center}
\end{figure}
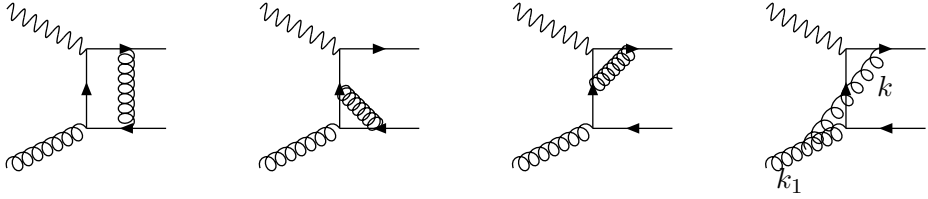
%--------------------------------------------
The order $\alpha_s$ corrections to the photon-gluon fusion process in 
Eq. (\ref{eqn:15}) have been calculated in \cite{lrsn1}.
Some of the virtual contributions to the Born reaction are shown in 
Fig. \ref{fig:4}.
The corresponding Feynman integrals, denoted by $\mu^{4-N} (2\pi)^{-N}
\int d^N k f(k,k_1)$, where $k_1$ is an external momentum,
reveal ultraviolet (UV), infrared (IR) and collinear (C)
divergences. They arise when the integration momentum $k$ takes the values
$ k \rightarrow \infty$ (UV), $k=0$ (IR) and $\vec k \Vert \vec k_1$ (C)
respectively. The former two divergences are very well known in quantum field
theory and we refer the reader to the textbooks. The collinear singularities
originate from the propagator $1/(k-k_1)^2$ present in $f(k,k_1)$. It becomes
singular when $k^2=k_1^2=0$ so that $(k-k_1)^2=-2\\
 \mid \vec k \mid\mid \vec k_1 \mid
(1 - \cos~\theta)$. The singularity then shows up at $\theta = 0$. The
appearance of IR- and C-divergences is due to the fact that we neglect 
confinement effects in the initial and final state while doing perturbation
theory. Here all external partons are put on-shell which is certainly not
possible if we would have applied nonperturbative methods. In this case the
partons are off-shell akin to the case when the particles constitute a bound 
state.
Fortunately there exist some theorems in quantum field theory which enable
us to remove the divergences mentioned above. This will be discussed below. 
For the
moment we have to find a way to define the Feynman integrals in which these
divergences occur. For that purpose we choose the method of $N$-dimensional
regularization which is the most suitable one since it preserves all Ward
identities characteristic of gauge field theories. Using this method the
divergences manifest themselves as pole terms of the type
$1 /\epsilon^k$ with $\epsilon = N -4$.\\
Besides the virtual gluon graphs in Fig. \ref{fig:4} there are also
contributions from the gluon bremsstrahlung diagrams in Fig. \ref{fig:5}.
This process is given by 
\begin{eqnarray}
\label{eqn:17}
\gamma^* + g \rightarrow c + \bar c + g\,.
\end{eqnarray}
Analogous to the one-loop Feynman integrals discussed above, IR and 
C divergences also show up in the phase space integrals corresponding to
process (\ref{eqn:17}). 
They arise when $k_2 \rightarrow 0$ (IR) and $\vec k_2 \Vert \vec k_1$
(C). The last condition originates from the propagator 
$f(k_1,k_2) \sim 1/(k_1-k_2)^2$ appearing in
the phase space integral $\mu^{4-N} \int d^{N-1} \vec k_2 (2E_2)^{-1} 
f(k_2,k_1)$.
Addition of the virtual corrections to process (\ref{eqn:15}) 
and the contributions from the bremsstrahlung
reaction in (\ref{eqn:17})) leads to a cancellation of all IR
singularities. This is called the Bloch-Nordsieck theorem \cite{blno}. 
Furthermore the Kinoshita-Lee-Nauenberg theorem \cite{kln} states that
collinear divergences which can be attributed to the final state cancel too.
This cancellation only happens when the final state is completely inclusive.
Notice that up to the order in $\alpha_s$ discussed above final state
collinear divergences are not present due to the heavy quark mass. 
The singularities
which remain are of UV and initial state collinear origin. The former are 
removed by mass and coupling constant renormalization. Mass renormalization is
performed by replacing the bare mass $\hat m$ in the lowest order coefficient 
function $H_{i,g}^{(1)}(z,Q^2,\hat m^2,\hat \alpha_s)$ via the substitution
%-------------------------------
%fig5
%----------------------------------------
\begin{eqnarray}
\label{eqn:18}
\hat m = m [ 1 + \hat a_s \Big( \frac{2}{\epsilon} \delta_0 + d_1\Big)] \quad
\mbox{with} \quad \hat a_s \equiv \frac{\hat\alpha_s}{4\pi}\,.
\end{eqnarray}
After the substitution one has to make an expansion around $\hat \alpha_s$ 
which involves taking derivatives of $H_{i,g}$ with respect to $m^2$. Further
$d_1$ is an arbitrary constant which is fixed by choosing the 
on-mass-shell scheme. After this renormalization the second order coefficient
function takes the following form.
\begin{eqnarray}
\label{eqn:19}
\hat H_{i,g}^{(2)}&=& \hat a_s \Big [ \{ \frac{1}{\epsilon_C} + \frac{1}{2} 
\ln(\frac{m^2}{\mu^2}) \} P_{gg}^{(0)} \otimes H_{i,g}^{(1)}
-\beta_0 \{ \frac{2}{\epsilon_{UV}} + \ln(\frac{m^2}{\mu^2}) \} H_{i,g}^{(1)}
\Big]
\nonumber\\
&& + H_{i,g}^{\rm finite,(2)}\,.
\end{eqnarray}
Here we have distinguished between the UV and the C-divergences
which are indicated by $1/\epsilon_{UV}$ and $1/\epsilon_C$ respectively.
Besides the higher order corrections to the photon-gluon fusion process
there also exists another $\alpha_s^2$ subprocess where the
photon couples to the charm quark. It is given by the Bethe-Heitler reaction
(see Fig. \ref{fig:6}).
%---------------------------------
%fig6
%--------------------------
Instead of a gluon we have now one of the light (anti-)quarks in the initial 
state and the reaction proceeds as follows
\begin{eqnarray}
\label{eqn:20}
\gamma^* + q(\bar q) \rightarrow c + \bar c + q(\bar q)\,.
\end{eqnarray}
In this process we only encounter collinear divergences and the 
bare heavy quark
coefficient function reads
\begin{eqnarray}
\label{eqn:21}
\hat H_{i,q}^{(2)}= \hat a_s \{ \frac{1}{\epsilon_C} + \frac{1}{2}
\ln(\frac{m^2}{\mu^2}) \} P_{gq}^{(0)} \otimes H_{i,g}^{(1)}
+ H_{i,q}^{\rm finite,(2)}\,.
\end{eqnarray}
Corrected up to order $\hat \alpha_s^2$ the structure function reads 
\begin{eqnarray}
\label{eqn:22}
F_{i,c} = e_c^2 [ \hat f_g^{\rm S} \otimes \{ \hat H_{i,g}^{(1)} 
+ \hat H_{i,g}^{(2)}\}
+\hat f_q^{\rm S} \otimes \hat H_{i,q}^{(2)} + \cdots ]\,.
\end{eqnarray}
Next we have to apply coupling constant renormalization in order to get rid of
the remaining UV divergence. This is achieved by replacing
\begin{eqnarray}
\label{eqn:23}
\hat \alpha_s = \alpha_s \Big[ 1 + a_s \beta_0 \Big ( \frac{2}{\epsilon} 
+ \ln(\frac{\mu_R^2}{\mu^2}) + b_1 \Big ) \Big]
\end{eqnarray}
in the coefficient function $\hat H_{i,g}^{(1)}(z,Q^2,m^2,\hat \alpha_s)$. In 
the above expression $\mu_R$ stands for the renormalization scale and $\alpha_s
\equiv \alpha_s(\mu_R^2)$. Further $\beta_0$ denotes the lowest order 
coefficient appearing in the beta-function which will be defined below.
The constant $b_1$ is scheme dependent. Here we will choose the 
$\overline{\rm MS}$-scheme so that $b_1=\gamma_E - \ln(4\pi)$ ($\gamma_E$ is 
the Euler constant).
Finally we have to perform mass factorization in 
order to get rid of the collinear divergence indicated by the pole term
$1/\epsilon_C$ in Eqs. (\ref{eqn:19}), (\ref{eqn:21}).
This is achieved by replacing the bare $\hat f_k$ by the renormalized 
parton density $f_k$. In general one has
\begin{eqnarray}
\label{eqn:24}
\hat f_l^{\rm S} = f_k^{\rm S}(\mu_F^2,\mu_R^2) \otimes \Big[ 
\delta_{kl} - a_s P_{kl}^{(0)} 
\Big (\frac{1}{\epsilon_C} + \frac{1}{2}\ln(\frac{\mu_F^2}{\mu^2}) + c_{kl}^
{(1)}\Big ) \Big] \,,
\end{eqnarray}
where $f_q^{\rm S}$ 
denotes the singlet combination of the light flavour densities
\begin{eqnarray}
\label{eqn:25}
f_q^{\rm S} &=& \sum_{k=1}^{n_f} [ f_k+f_{\bar k}]\,.
\end{eqnarray}
For $n_f=3$, the index $k$ runs over ($u,d,s$). Further
$\mu_F$ stands for the factorization scale and $P_{kl}^{(0)}$ ($k,l=q,g$)
denote the lowest order DGLAP splitting functions. 
The constants $c_{kl}^{(1)}$ are scheme dependent. Here we will choose the
$\overline{\rm MS}$-scheme which implies that $c_{kl}=\gamma_E - \ln(4\pi)$.
One can also perform mass factorization on the level of the heavy quark
coefficient functions. In this case one gets the equation

\begin{eqnarray}
\label{eqn:26}
&& \hat H_{i,k}(Q^2,m^2,\alpha_s,\mu_R^2,\epsilon_C,\mu^2)= 
\nonumber\\
&& \Gamma_{lk} (\alpha_s,
\mu_R^2,\mu_F^2,\epsilon_C,\mu^2) \otimes 
H_{i,l}(Q^2,m^2,\alpha_s,\mu_R^2,\mu_F^2)\,,
\end{eqnarray}
where $\Gamma_{kl}$ denotes the transition function given by
\begin{eqnarray}
\label{eqn:27}
\Gamma_{kl}= \delta_{kl} + a_s P_{kl}^{(0)} \Big ( \frac{1}{\epsilon_C} 
+ \frac{1}{2}\ln(\frac{\mu_F^2}{\mu^2}) + c_{kl}^{(1)} \Big) + \cdots\,.
\end{eqnarray}
{}From Eq. (\ref{eqn:24}) we infer that
\begin{eqnarray}
\label{eqn:28}
 f_k^{\rm S} (\mu_R^2,\mu_F^2) 
= \Gamma_{kl} (\alpha_s,\mu_R^2,\mu_F^2,\epsilon_C,
\mu^2)\otimes \hat f_l^{\rm S}\,. 
\end{eqnarray}
In order to get finite coefficient functions up to order $\alpha_s^2$
we need the following transition functions
\begin{eqnarray}
\label{eqn:29}
\Gamma_{qq} &=& 1 + O(\alpha_s)\,,
\nonumber\\
\Gamma_{qg} &=& O(\alpha_s)\,,
\nonumber\\
\Gamma_{gq}&=&a_s P_{gq}^{(0)} \Big ( \frac{1}{\epsilon_C} + \frac{1}{2}
\ln(\frac{\mu_F^2}{\mu^2}) + c_{gq}^{(1)} \Big) + \cdots\,,
\nonumber\\
\Gamma_{gg}&=&  1 + a_s  P_{gg}^{(0)} \Big ( \frac{1}{\epsilon_C} + \frac{1}{2}
\ln(\frac{\mu_F^2}{\mu^2}) + c_{gg}^{(1)} \Big) + \cdots\,.
\end{eqnarray}
%-----------------------------------
%fig7
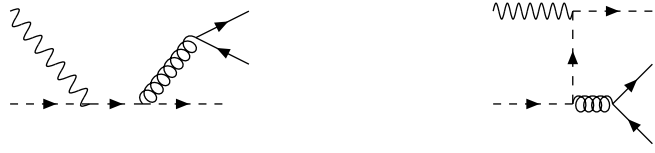
\begin{figure}
\begin{center}
  \begin{picture}(90,50)
  \DashArrowLine(0,15)(30,15){3}
  \DashArrowLine(30,15)(50,15){3}
  \DashArrowLine(50,15)(80,15){3}
  \ArrowLine(70,40)(90,50)
  \ArrowLine(90,30)(70,40)
  \Gluon(50,15)(70,40){3}{7}
  \Photon(0,50)(30,15){3}{7}
  \end{picture}
\hspace*{3cm}
%  \hfill
  \begin{picture}(60,50)
  \DashArrowLine(0,15)(30,15){3}
  \DashArrowLine(30,15)(30,50){3}
  \DashArrowLine(30,50)(60,50){3}
  \Gluon(30,15)(45,15){3}{4}
  \ArrowLine(45,15)(60,30)
  \ArrowLine(60,0)(45,15)
  \Photon(0,50)(30,50){3}{7}
\end{picture}
\vspace*{5mm}
  \caption{The Compton process $\gamma^* + q \rightarrow
           c + \bar c + q$ contributing to the coefficient functions
           $L_{i,q}^{(2)}$.}
  \label{fig:7}
\end{center}
\end{figure}
%-------------------------------
Notice that we have suppressed the $z$-dependence of the functions 
$P_{kl}^{(0)}$, $c_{kl}^{(1)}$ and $1 \equiv \delta(1-z)$.
After renormalization
and mass factorization the charm component of the structure function reads
\begin{eqnarray}
\label{eqn:30}
F_{i,c}= e_c^2 [ f_g \otimes \{ H_{i,g}^{(1)} + H_{i,g}^{(2)} \} + f_q \otimes
H_{i,q}^{(2)} ]\,,
\end{eqnarray}
where $H_{i,g}^{(1)}$ is obtained from $\hat H_{i,g}^{(1)}$ by the substitution
: $\hat m \rightarrow m$ and $\hat \alpha_s \rightarrow \alpha_s$. Further
we introduce the shorthand notations
\begin{eqnarray}
\label{eqn:31}
H_{i,g}^{(2)}&=&  H_{i,g}^{\rm finite,(2)} +
a_s \Big ( \frac{1}{2} \ln(\frac{m^2}{\mu_F^2}) P_{gg}^{(0)} \otimes
H_{i,g}^{(1)} 
%\nonumber\\&& 
-\beta_0 H_{i,g}^{(1)} \ln(\frac{m^2}{\mu_R^2}) \Big)\,,
\end{eqnarray}
\begin{eqnarray}
\label{eqn:32}
H_{i,q}^{(2)}=  H_{i,q}^{\rm finite,(2)} + a_s \Big (\frac{1}{2}
\ln(\frac{m^2}{\mu_F^2}) P_{gq}^{(0)} \otimes H_{i,g}^{(1)} \Big )\,. 
\end{eqnarray}
Besides the reactions discussed above where the photon interacts with the
heavy quark the former can also couple to the light (anti-)quark. This happens 
for the first time in order $\alpha_s^2$. In this order charm production
proceeds via the Compton process (Fig. \ref{fig:7})) 
which is given by Eq. \ref{eqn:20} where now the
photon couples to the light (anti-)quark. The corresponding coefficient 
function is denoted by $L_{2,q}^{\rm NS}$ which in order $\alpha_s^2$
is equal to $L_{2,q}^{\rm S}$. The
structure function $F_{i,c}$ gets the contribution
\begin{eqnarray}
\label{eqn:33}
F_{i,c}= \frac{1}{n_f}\sum_{k=1}^{n_f} e_k^2 \Big [ f_q^{\rm S} 
\otimes L_{i,q}^{\rm S}
+ n_f f_k^{\rm NS} \otimes L_{i,q}^{\rm NS} \Big ]\,.
\end{eqnarray}
The non-singlet combination of the quark densities is defined by
\begin{eqnarray}
\label{eqn:34}
f_k^{\rm NS} &=& f_k + f_{\bar k}
-\frac{1}{n_f}f_q^{\rm S} \hspace*{1cm} \mbox{($k=u,d,s$)}\,.
\end{eqnarray}
%---------------------
%fig8
\begin{figure}
 \begin{center}
  {\unitlength1cm
  \epsfig{file=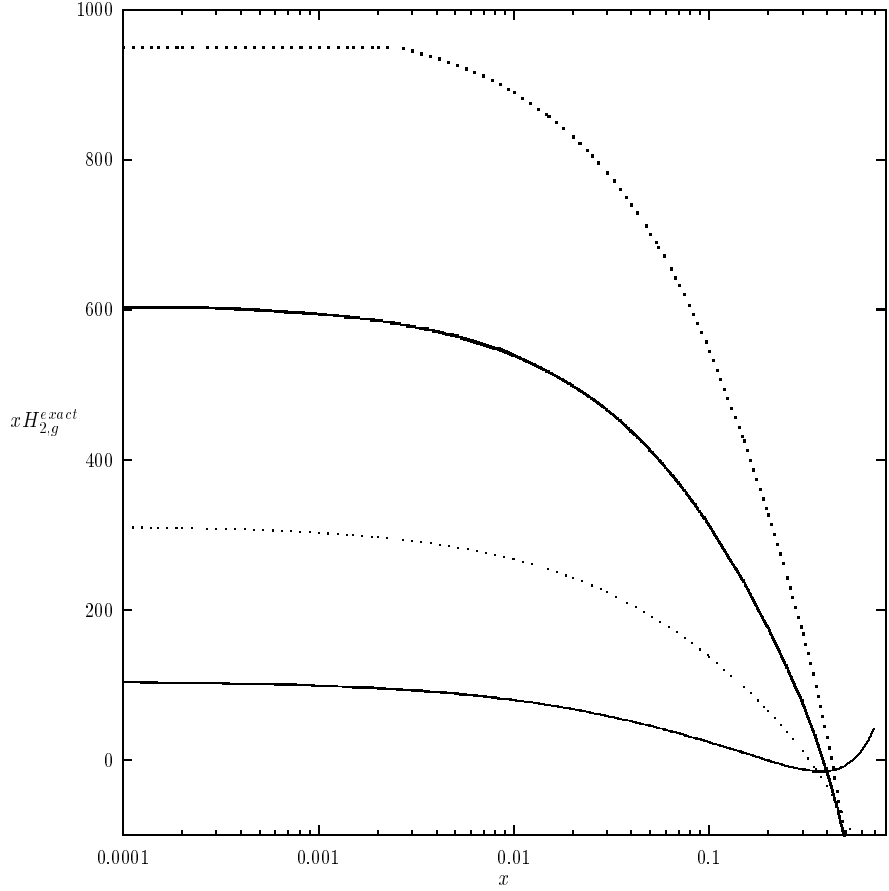,bbllx=52pt,bblly=272pt,bburx=536pt,bbury=710pt,%
    height=12cm,width=12cm,clip=,angle=0}
     }
\end{center}
\caption[]{The function $H_{2,g}(x,Q^2,m_c^2)$ for $Q^2=10$ 
(lower solid line), $Q^2=100$ (lower dotted line), 
$Q^2=10^3$ (upper solid line),
$Q^2=10^4$ (upper dotted line). All units are in $({\rm GeV/c})^2$}. 
  \label{fig:8}
\end{figure}
%-----------------------
Up to second order no collinear divergences appear in this type of processes.
However in higher order they do appear. Moreover there are reactions with a
gluon in the initial state leading to the heavy quark coefficient function
$L_{i,g}^{\rm S}$. Together with $L_{i,q}^{\rm S}$ 
they satisfy the same mass factorization
relations as presented for $H_{i,k}^{\rm S}$ in Eq. (\ref{eqn:26}).
For the non-singlet part a simpler relation holds (no mixing) which is given by
\begin{eqnarray}
\label{eqn:35}
\hat L_{i,q}^{\rm NS} &=& \Gamma_{qq}^{\rm NS} L_{i,q}^{\rm NS}\,.
\end{eqnarray}
Likewise for the non-singlet parton density we have
\begin{eqnarray}
\label{eqn:36}
f_k^{\rm NS} &=& \Gamma_{qq}^{\rm NS} \hat f_k^{\rm NS}\,.
\end{eqnarray}
The reason that the transition functions $\Gamma_{kl}$ are the same for
$H_{i,k}$ and $L_{i,k}$ follows from the universality of collinear divergences.
It means that the latter are process independent. The same residues, 
represented by the DGLAP functions $P_{kl}$, are also found while calculating
corrections to other hard processes differing from 
heavy flavour production. One of the
most important consequences is that 
the finite parton densities also become universal
after mass factorization so that one can use them as input for other 
processes to yield absolute predictions.
%----------------------------
%fig9
\begin{figure}
 \begin{center}
 {\unitlength1cm
    \epsfig{file=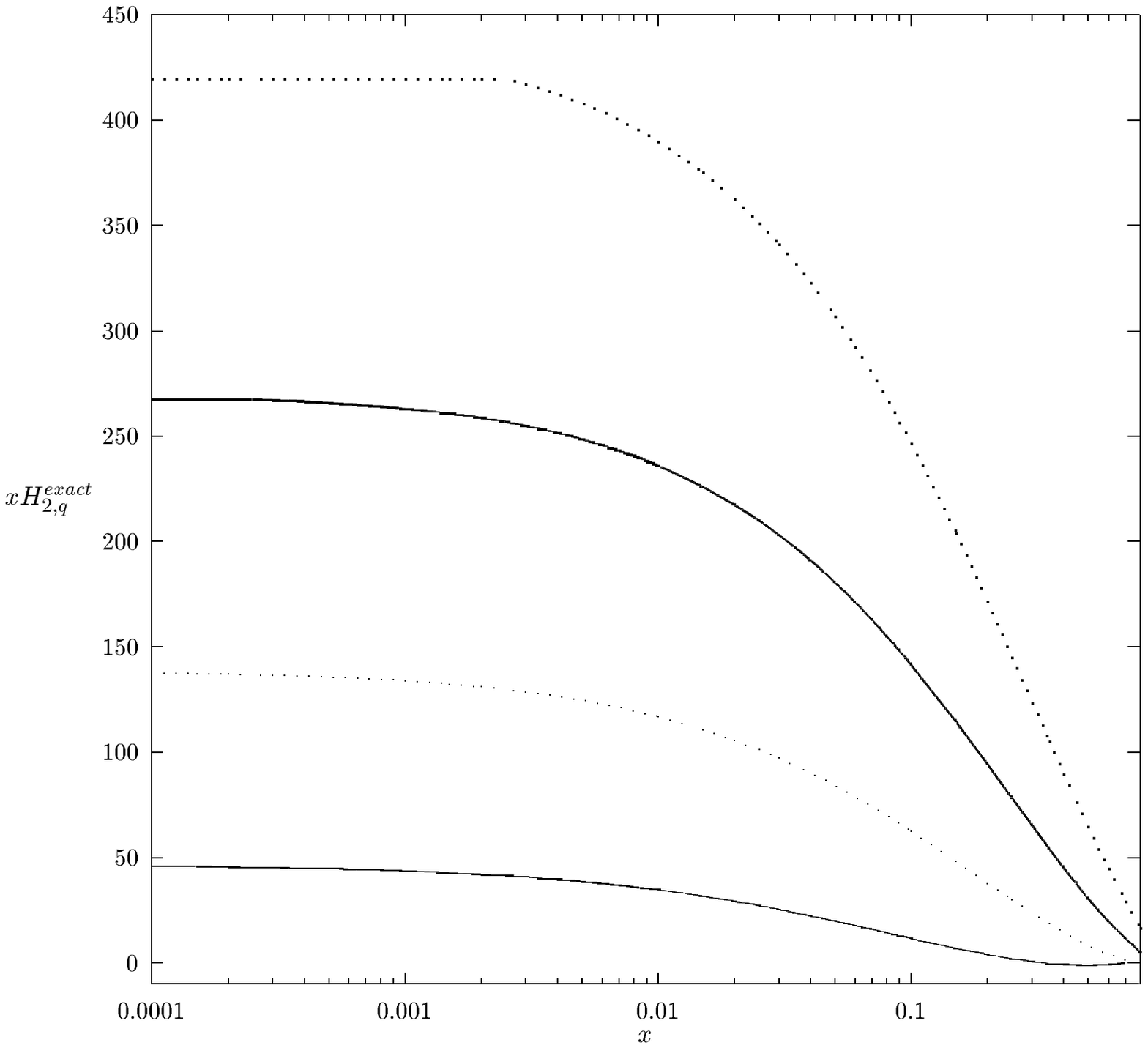,bbllx=52pt,bblly=272pt,bburx=536pt,bbury=710pt,%
      height=12cm,width=12cm,clip=,angle=0}
     }
 \end{center}
\caption[]{Same as fig. 8 for $H_{2,q}(x,Q^2,m_c^2)$.}
  \label{fig:9}
\end{figure}
%-----------------------------------------------
Collecting all contributions the charm component of the deep inelastic 
structure function for the proton reads 
\begin{eqnarray}
\label{eqn:37}
 F_{i,c}(n_f,Q^2,m^2) &=& 
 \frac{1}{n_f} \sum_{k=1}^{n_f} e_k^2 
\Big [ f_q^{\rm S}(n_f,\mu^2) \otimes L_{i,q}^{\rm S}(n_f,Q^2,m^2,\mu^2)
\nonumber\\
&& + f_g^{\rm S}(n_f,\mu^2) \otimes L_{i,g}^{\rm S}(n_f,Q^2,m^2,\mu^2)
\nonumber\\
&& + n_f f_k^{\rm NS}(n_f,\mu^2) \otimes L_{i,q}^{\rm NS }
(n_f,Q^2,m^2,\mu^2) \Big ]
\nonumber\\
&& + e_c^2 \Big [ f_q^{\rm S}(n_f,\mu^2) 
\otimes H_{i,q}^{\rm PS}(n_f,Q^2,m^2,\mu^2) 
\nonumber\\
&& + f_g^{\rm S}(n_f,\mu^2) \otimes H_{i,g}^{\rm S}(n_f,Q^2,m^2,\mu^2) 
\Big ] \,,
\end{eqnarray}
where now all quantities are finite. Notice that contrary to $L_{i,k}$
the functions $H_{i,k}$ are purely singlet (PS) only. This automatically holds
for $H_{i,g}$. In the case of $k=q$ we have added the superscript PS
because $H_{i,k}^{\rm NS}=0$ (no intrinsic charm !!!). As we have shown above
the coefficient functions and the parton densities depend on two different 
scales $\mu_R$ and $\mu_F$. In order to simplify the renormalization group 
equations (RGE's) below we will set them equal i.e. $\mu_R=\mu_F=\mu$.
Let us first define the total derivative with respect to $\mu$.
\begin{eqnarray}
\label{eqn:38}
D \equiv \mu \frac{d}{d\mu} = \mu \frac{\partial}{\partial \mu} 
+ \beta(\alpha_s) \frac{\partial}{\partial \alpha_s} \,,
\end{eqnarray}
where $\beta(\alpha_s)$ denotes the beta-function which can be written
as a series expansion in the strong coupling constant
\begin{eqnarray}
\label{eqn:39}
\beta(\alpha_s) = -2\beta_0 \alpha_s^2 - 2 \beta_1 \alpha_s^3 + \cdots \,.
\end{eqnarray}
One can show that the parton densities satisfy the following RGE's
\begin{eqnarray}
\label{eqn:40}
D f_q^{\rm NS} = P_{qq}^{\rm NS} \otimes f_q^{\rm NS} 
\quad , \quad D f_k^{\rm S} = P_{kl}^{\rm S} 
\otimes f_l^{\rm S} \,.
\end{eqnarray}
The heavy quark coefficient functions satisfy the RGE's
\begin{eqnarray}
\label{eqn:41}
D L_{i,q}^{\rm NS} &=& - P_{qq}^{\rm NS}\otimes L_{i,q}^{\rm NS} \quad
D L_{i,k}^{\rm S} = - P_{lk}^{\rm S}\otimes L_{i,l}^{\rm S} 
\nonumber\\
D H_{i,k}^{\rm S} &=& - P_{lk}^{\rm S}\otimes H_{i,l}^{\rm S} \,.
\end{eqnarray}
The DGLAP splitting functions can be expanded in the coupling constant
\begin{eqnarray}
\label{eqn:42}
P_{kl}= a_s P_{kl}^{(0)} + a_s^2 P_{kl}^{(1)} + \cdots \,.
\end{eqnarray}
%---------------------------------
%fig10
\begin{figure}
 \begin{center}
  {\unitlength1cm
   \epsfig{file=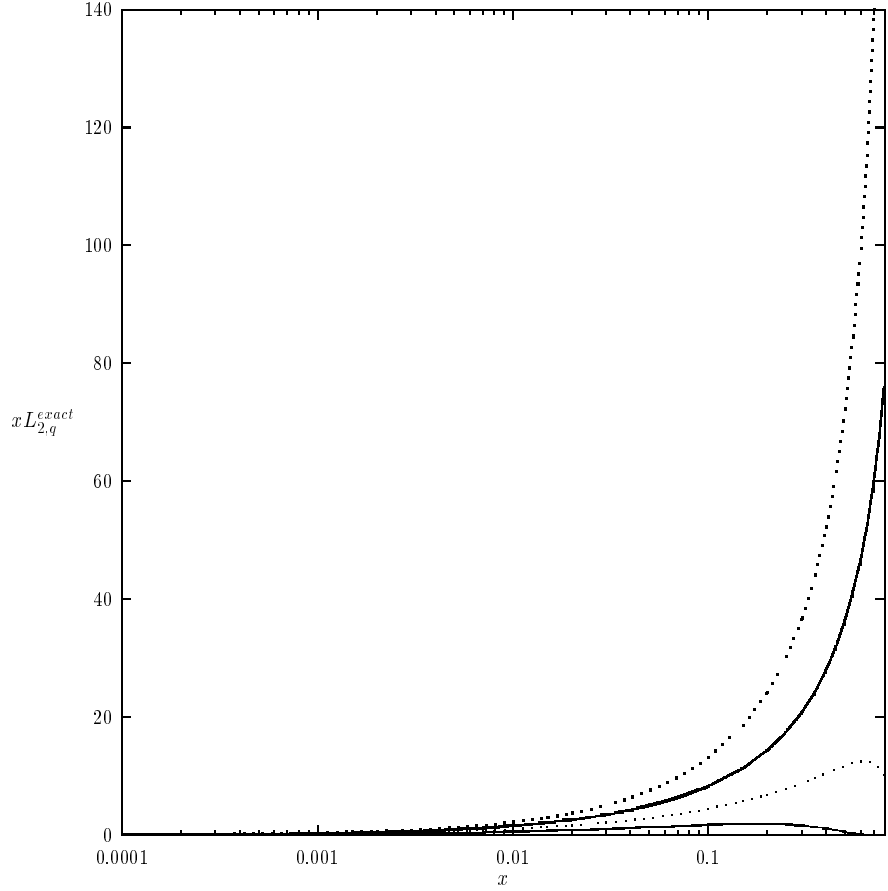,bbllx=52pt,bblly=272pt,bburx=536pt,bbury=710pt,%
       height=12cm,width=12cm,clip=,angle=0}
    }
  \end{center}
 \caption[]{Same as fig. 8 for $L_{2,q}(x,Q^2,m_c^2)$.}
  \label{fig:10}
\end{figure}
%--------------------------------------------
Using the above equations one can now show that 
$F_{i,c}$ is a renormalization 
group invariant i.e. $D~F_{i,c}=0$. It means that it is a physical quantity
which is independent of the scale $\mu$ and the chosen scheme. (Notice that in 
these lectures we have adopted the $\overline{\rm MS}$-scheme for
the renormalization of the coupling constant and for mass factorization).\\
The order $\alpha_s^2$ contributions to the heavy quark coefficient functions
$H_{i,k}$ and $L_{i,k}$ have been computed in \cite{lrsn1}.
 The expressions are so
complicated that it is impossible to give analytic results. They are buried
in long programs containing the numerical computation of
two dimensional integrals. In order to make these coefficient functions more 
amenable for phenomenological applications they are tabulated \cite{rsn}
in the form
of a two dimensional array in the variables 
\begin{eqnarray}
\label{eqn:43}
\eta = \frac{1-z}{4z} \xi - 1 \quad , \quad  \xi=\frac{Q^2}{m^2}\,.
\end{eqnarray}
%--------------------------
%fig11
\begin{figure}
\begin{center}
  \begin{picture}(130,60)
  \Gluon(0,0)(60,0){3}{7}
  \ArrowLine(60,30)(30,30)
  \ArrowLine(30,30)(30,50)
  \ArrowLine(30,50)(60,50)
  \Gluon(30,0)(30,30){3}{4}
  \Photon(0,60)(30,50){3}{7}
  \DashLine(65,60)(65,0){3}
  \Gluon(70,0)(130,0){3}{7}
  \ArrowLine(70,30)(100,30)
  \ArrowLine(100,30)(100,50)
  \ArrowLine(100,50)(70,50)
  \Gluon(100,0)(100,30){3}{4}
  \Photon(130,60)(100,50){3}{7}
  \end{picture}
\hspace*{1cm}
  \begin{picture}(130,60)
  \Gluon(0,0)(30,30){3}{7}
  \ArrowLine(60,30)(30,30)
  \ArrowLine(30,30)(30,50)
  \ArrowLine(30,50)(60,50)
  \Gluon(15,15)(45,30){3}{4}
  \Photon(0,60)(30,50){3}{7}
  \DashLine(65,60)(65,0){3}
  \Gluon(100,30)(130,0){3}{7}
  \ArrowLine(70,30)(100,30)
  \ArrowLine(100,30)(100,50)
  \ArrowLine(100,50)(70,50)
  \Photon(130,60)(100,50){3}{7}
  \end{picture}
\vspace*{5mm}
  \caption{Soft gluon contributions to the coefficient functions
           $H_{i,g}^{(2)}$.}
  \label{fig:11}
\end{center}
\end{figure}
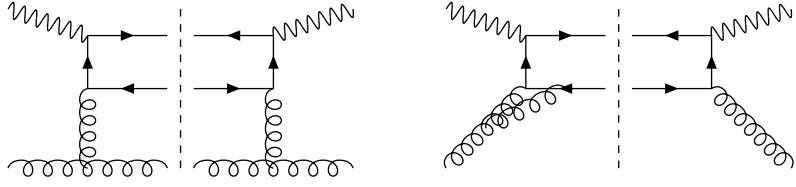
%-------------------------------
The first variable is chosen in such a way that the threshold region 
$\eta=(s-4m^2)(4m^2)^{-1} \sim 0$ is exposed in a clearer way. Notice that
this region dominates the integrand of the structure function (see \cite{vogt})
\begin{eqnarray}
\label{eqn:44}
F_{i,c}(x,Q^2,m^2) \sim \int_x^{z_{\rm max}} \frac{dz}{z}
 \, f_k(x/z) H_{i,k}(z,Q^2,m^2)\,.
\end{eqnarray}
In Figs. \ref{fig:8}, \ref{fig:9}, \ref{fig:10} we have plotted the coefficient
functions $H_{2,g}$, $H_{2,q}$ and $L_{2,q}$ respectively. Here we have chosen
$m=m_c=1.5~{\rm GeV/c}$. From these plots
we infer that $H_{2,g}$ is much larger than the
two other heavy quark coefficient functions so that the photon-gluon fusion
process constitutes the bulk of the radiative corrections. Since this process
only depends on the gluon density $f_g^{\rm S}(z,\mu^2)$ 
charm electroproduction yields 
a measurement of the latter density, in particular at small $z$.
The coefficient functions reveal large corrections which occur in two regions
given by
\begin{itemize}
\item[1] \underline{Threshold regime:}
 $s \sim 4m^2$ or $z \sim z_{\rm max} = Q^2/(Q^2+4m^2)$\\[3mm]
Here we have two types of large corrections. The first one is due to
soft gluon bremsstrahlung. As we have discussed before the real gluon
and the virtual gluon corrections show IR divergences when $k_2 \rightarrow 0$.
In the real gluon case we have a phase space factor and the corresponding
coefficient function behaves like
\begin{eqnarray}
\label{eqn:45}
H_{i,g}^{(2),{\rm REAL}} &\sim & \frac{1}{\epsilon_{\rm IR}} 
\Big ( \frac{s-4m^2}{4m^2} 
\Big )^{\epsilon_{{\rm IR}/2}} H_{i,g}^{(1)}\,.
\end{eqnarray}
On the other hand the virtual contribution equals
\begin{eqnarray}
\label{eqn:46}
H_{i,g}^{(2),{\rm VIRT}} &\sim & - \frac{1}{\epsilon_{\rm IR}} H_{i,g}^{(1)}\,.
\end{eqnarray}
Addition of the two expressions above leaves the finite result
\begin{eqnarray}
\label{eqn:47}
H_{i,g}^{(2)} \sim \ln\Big(\frac{s-4m^2}{4m^2}\Big) \sigma_{i,g}^{(1)}\,,
\end{eqnarray}
which however blows up when $s \rightarrow 4m^2$. The above behaviour is
typical for QED. In QCD we have a C-divergence 
in addition to the IR divergence.
Hence after having removed the IR and C-singularities the power  
of the logarithms will be doubled. The final result in QCD shows the 
following behaviour
\begin{eqnarray}
\label{eqn:48}
H_{i,g}^{(n)} \sim \alpha_s^n \ln^{2(n-1)}
\Big( \frac{s-4m^2}{4m^2} \Big) H_{i,g}^{(1)}\,.
\end{eqnarray}
The above logarithms can be summed in all orders of perturbation theory by
exponentiating them (see \cite{lsn}). The effect of these logarithms is shown
in Fig. \ref{fig:8} where we plotted $H_{2,g}^{(2)}$. In the threshold region
where $x$ is large and $Q^2$ is small one observes a rise in the coefficient
function.\\ 
%----------------------
%fig12
\begin{figure}
\begin{center}
  \begin{picture}(130,60)
  \Gluon(0,0)(30,15){3}{7}
  \ArrowLine(60,15)(30,15)
  \ArrowLine(30,15)(30,45)
  \ArrowLine(30,45)(60,45)
  \Gluon(45,45)(45,15){3}{7}
  \Photon(0,60)(30,45){3}{7}
  \DashLine(65,60)(65,0){3}
  \Gluon(130,0)(100,15){3}{7}
  \ArrowLine(70,15)(100,15)
  \ArrowLine(100,15)(100,45)
  \ArrowLine(100,45)(70,45)
  \Photon(130,60)(100,45){3}{7}
  \end{picture}
\vspace*{5mm}
  \caption{The Coulomb singularity appearing in the box graph of the process
           $\gamma^* + g \rightarrow
           c + \bar c$.}
  \label{fig:12}
\end{center}
\end{figure}
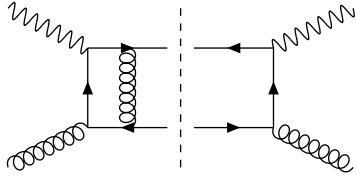
%-------------------------------
Another large correction near threshold
can be traced back to the Coulomb singularity appearing in the virtual
contribution in Fig. \ref{fig:12}. One obtains the following expression 
\begin{eqnarray}
\label{eqn:49}
\mid M_g^{(2),{\rm VIRT}} \mid ^2  \sim \frac{\pi^2}{\sqrt(s-4m^2)}
\quad \rightarrow \quad H_{i,g}^{(2),{\rm VIRT}} \sim  \frac{\pi^2}{m^2}\,.
\end{eqnarray}
This singularity is enhanced by multiple gluon
exchanges between the heavy quark lines. It arises because we apply 
perturbation theory. The correct way to deal with these exchanges is to
apply nonperturbative methods. In this case the nonperturbative result
shows a singularity at $\alpha_s=0$ which implies that one cannot make
an expansion around this point. In practice the effect of the Coulomb
singularity is so small that one can hardly see it in the heavy quark
coefficient functions.
 item[2] \underline{Asymptotic regime:}
 $s \gg m^2$ or $z \rightarrow 0$.\\[3mm]
The large corrections in this region are due to the exchange of multiple
soft gluons in the t-channel. 
The corresponding coefficient functions have the form
\begin{eqnarray}
\label{eqn:50}
H_{i,k}^{(n)}(z,Q^2,m^2) 
\mathop{\sim}\limits_{\vphantom{\frac{A}{A}} z \rightarrow 0}
\frac{1}{z} \ln^{n-2} (z) h(Q^2,m^2)
\quad \mbox{for $n \geq 2;~~ k=q,g$}\,.
\end{eqnarray}
For $n=2$ see Fig. \ref{fig:13}.
Like in the previous cases these large corrections can be resummed (see 
\cite{cch}).
The soft gluon exchange mechanism has a huge effect on the heavy quark
coefficient functions. It is responsible for the large plateau in the small $x$
region of the functions $H_{2,g}^{(2)}$ (Fig. \ref{fig:8}) and 
$H_{2,q}^{(2)}$ (Fig. \ref{fig:9}).
This plateau rises as $Q^2$ increases. However it turns out that this
production mechanism is not so important for the behaviour of $F_{i,c}$ 
in Eq. (\ref{eqn:44})
at small $x$ since the the main contribution to the integral comes from
large $z$ rather than small $z$.
\end{itemize}
\section{Asymptotic heavy quark coefficient functions}
\begin{figure}
\begin{center}
  \begin{picture}(130,70)
  \Gluon(0,0)(30,15){3}{7}
  \Photon(0,60)(30,45){3}{7}
  \ArrowLine(90,15)(30,15)
  \ArrowLine(30,15)(30,45)
  \ArrowLine(30,45)(90,45)
  \ArrowLine(90,45)(90,15)
  \Gluon(120,0)(90,15){3}{7}
  \Photon(120,60)(90,45){3}{7}
    \Text(0,35)[t]{Im}
    \Text(15,70)[t]{$\downarrow q$}
    \Text(105,70)[t]{$\uparrow q$}
    \Text(15,0)[t]{$\uparrow p$}
    \Text(105,0)[t]{$\downarrow p$}
    \Text(60,60)[t]{$k+q$}
    \Text(23,35)[t]{$k$}
    \Text(97,35)[t]{$k$}
    \Text(60,10)[t]{$k-p$}
  \end{picture}
\hspace*{1cm}
  \begin{picture}(130,70)
  \Gluon(0,0)(30,15){3}{7}
  \ArrowLine(60,15)(30,15)
  \ArrowLine(30,15)(30,45)
  \ArrowLine(30,45)(60,45)
  \Photon(0,60)(30,45){3}{7}
  \DashLine(65,60)(65,0){3}
  \Gluon(130,0)(100,15){3}{7}
  \ArrowLine(70,15)(100,15)
  \ArrowLine(100,15)(100,45)
  \ArrowLine(100,45)(70,45)
  \Photon(130,60)(100,45){3}{7}
    \Text(0,30)[t]{=}
  \end{picture}
\vspace*{5mm}
  \caption{Forward Compton scattering : $\gamma^* + g \rightarrow \gamma^*
           + g$ and the optical theorem}
  \label{fig:14}
\end{center}
\end{figure}
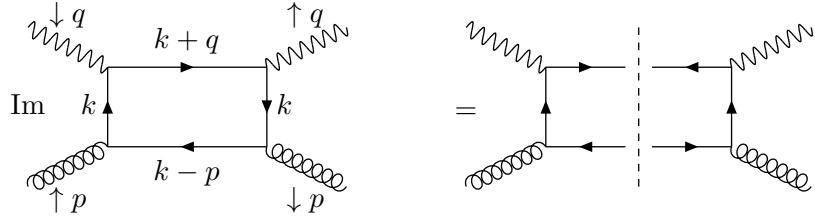

As we mentioned in the previous section it is very hard to get 
analytical expressions for the second order heavy quark coefficient functions
except for $L_{i,q}$ ($i=2,L$) which are published in \cite{bmsmn}. However when
$Q^2 \gg m^2$ it is possible to obtain analytical expressions. This is very
useful because the latter serve as a check on the exact calculations carried 
out in \cite{lrsn1},\cite{rsn}. Furthermore it turns out that for charm
production the asymptotic heavy quark coefficient functions give an equally
good description as the exact ones when $Q^2>20 \, ({\rm GeV/c})^2$ 
and $x<0.01$.
%---------------------------
%fig13
\begin{figure}
\begin{center}
  \begin{picture}(130,60)
  \Gluon(0,0)(60,0){3}{7}
  \ArrowLine(60,30)(30,30)
  \ArrowLine(30,30)(30,50)
  \ArrowLine(30,50)(60,50)
  \Gluon(30,0)(30,30){3}{4}
  \Photon(0,60)(30,50){3}{7}
  \DashLine(65,60)(65,0){3}
  \Gluon(70,0)(130,0){3}{7}
  \ArrowLine(70,30)(100,30)
  \ArrowLine(100,30)(100,50)
  \ArrowLine(100,50)(70,50)
  \Gluon(100,0)(100,30){3}{4}
  \Photon(130,60)(100,50){3}{7}
  \end{picture}
\hspace*{1cm}
  \begin{picture}(130,60)
  \DashArrowLine(0,0)(30,0){2}
  \DashArrowLine(30,0)(60,0){2}
  \ArrowLine(60,30)(30,30)
  \ArrowLine(30,30)(30,50)
  \ArrowLine(30,50)(60,50)
  \Gluon(30,0)(30,30){3}{4}
  \Photon(0,60)(30,50){3}{7}
  \DashLine(65,60)(65,0){3}
  \DashArrowLine(130,0)(100,0){2}
  \DashArrowLine(100,0)(70,0){2}
  \ArrowLine(70,30)(100,30)
  \ArrowLine(100,30)(100,50)
  \ArrowLine(100,50)(70,50)
  \Gluon(100,0)(100,30){3}{4}
  \Photon(130,60)(100,50){3}{7}
  \end{picture}
\vspace*{5mm}
  \caption{Soft gluon exchanges in the t-channel of the processes:
           $\gamma^* + g \rightarrow c + \bar c + g$ and
           $\gamma^* + q \rightarrow c + \bar c + q$.}
  \label{fig:13}
\end{center}
\end{figure}
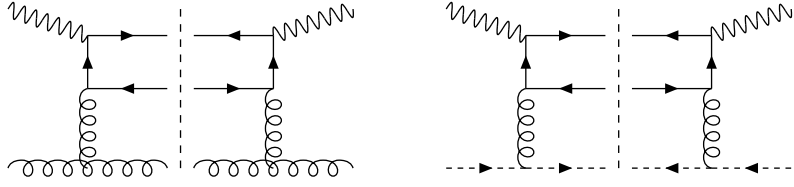
%-----------------------------------
Finally the asymptotic expressions can be also used for the charm component of
the structure function (Eq. \ref{eqn:37}) in the so called variable flavour
number scheme (VFNS) discussed in the next section.
In order to compute the asymptotic heavy quark coefficient functions one can 
proceed in two ways.
\begin{itemize} 
\item[1] \underline{Standard method}\\[3mm]
In this case one evaluates all Feynman and phase space integrals for 
$Q^2 \gg m^2$.
This method is still elaborate and one has to be careful with terms 
in these integrals which are proportional to $(m^2)^j$ because they
survive in the limit indicated above.
\item[2] \underline{Techniques of asymptotic expansions}\\[3mm]
Here we can use the operator product expansion (OPE) which is very elegant.
As we will explain by presenting some examples one can compute the asymptotic
expressions for $H_{i,k}$ and $L_{i,k}$ provided we know the massless
parton coefficient functions denoted by ${\cal C}_{i,k}$ and the heavy
quark operator matrix elements (OME's) given by $A_{ck}$ and $A_{kl,c}$.
($k,l=q,g$). Fortunately the functions ${\cal C}_{i,k}$ are already known
up to order $\alpha_s^2$ for some time \cite{zn}. Recently the OME's
have also been computed up the same order (see \cite{bmsmn},\cite{bmsn})
so that one can obtain the heavy quark coefficient
functions without doing too much work.
\end{itemize}
Adopting the second method in the subsequent part of this section it is very
easy to show that one obtains the following result
in lowest order 
\begin{eqnarray}
\label{eqn:51}
H_{2,g}^{\rm ASYMP,(1)}(z,Q^2,m^2) = a_s [ P_{qg}^{(0)}(z) 
\ln\Big(\frac{Q^2}{m^2}\Big) 
+ h_{2,g}^{(1)}(z)]\,.
\end{eqnarray}
The above equation can be easily inferred from the literature \cite{wit},
\cite{lrsn1} by taking the
limit $Q^2 \gg m^2$ of the exact formula. 
The latter can be also obtained in the
limit $m \rightarrow 0$. In this case the charm mass acts as a regulator
for the C-divergence which is indicated by its residue represented by the
DGLAP splitting function $P_{qg}$ \cite{dglap}. 
Because of the analogy between $\ln m^2$
and the pole term $1/\epsilon_C$ which represents the collinear divergence 
in the case of massless partons treated in the last section we can now
apply mass factorization to remove them from the asymptotic heavy quark
coefficient functions. In lowest order this is very simple and we obtain
\begin{eqnarray}
\label{eqn:52}
H_{2,g}^{\rm ASYMP,(1)}(z,Q^2,m^2) = {\cal C}_{2,g}^{(1)}
\Big(z,\frac{Q^2}{\mu^2}\Big) 
+ A_{cg}^{(1)}\Big(z,\frac{\mu^2}{m^2}\Big)\,, 
\end{eqnarray}
with
\begin{eqnarray}
\label{eqn:53}
{\cal C}_{2,g}^{(1)}\Big(z,\frac{Q^2}{\mu^2}\Big)= a_s[P_{qg}^{(0)}(z) 
\ln\Big(\frac{Q^2}{\mu^2}\Big) 
+ c_{2,g}^{(1)}(z)]\,,
\end{eqnarray}
and
\begin{eqnarray}
\label{eqn:54}
A_{cg}^{(1)}\Big(z,\frac{\mu^2}{m^2}\Big)= a_s 
[ P_{qg}^{(0)}(z) \ln\Big(\frac{\mu^2}{m^2}\Big) 
+ a_{cg}^{(1)}(z)]\,.
\end{eqnarray}
Here we want to stress the analogy between the quantities treated in the
previous and the present section. They are
\[
  \mbox{section 3} \hspace*{3cm} \mbox{section 4}
\]
\begin{eqnarray}
\label{eqn:55}
&& \hat H_{i,k}(z,Q^2,m^2,\epsilon_C) \hspace*{3cm}  
H_{i,k}^{\rm ASYMP}(z,Q^2,m^2)
\nonumber\\
&& H_{i,k}(z,Q^2,m^2,\mu^2)  \hspace*{3cm}  
{\cal C}_{i,k}\Big(z,\frac{Q^2}{\mu^2}\Big)
\nonumber\\
&& \Gamma_{kl}(z,\mu^2,\epsilon_C) \hspace*{40mm} 
A_{ck}\Big(z,\frac{\mu^2}{m^2}\Big)
\,.
\end{eqnarray}
In particular we want to emphasize the similar role played by the transition 
functions $\Gamma_{kl}$ and the OME's $A_{ck}$. Both of them absorb the 
collinear divergences from the original quantities. 
We will now show that in lowest order
the quantity $A_{cg}^{(1)}$ really stands for the heavy quark OME.\\
Squaring the amplitude corresponding to the Feynman graphs in Fig. \ref{fig:1}
and integrating over the final particle 
phase space one can apply the optical theorem
which states that
\begin{eqnarray}
\label{eqn:56}
H_{i,k}(s,Q^2,m^2) = \mbox{Im} \,  T_{i,k}(s,Q^2,m^2)\,,
\end{eqnarray}
where $z=s/(s+Q^2)$. Here $T_{i,k}$ stands for the forward Compton scattering
amplitude (Fig \ref{fig:14})
and the discontinuity denoted by Im is taken over the s-channel.
If we define $\nu=p\cdot q=(s-m^2+Q^2)/2$ one can write 
an unsubtracted dispersion
relation which can be expressed as follows
\begin{eqnarray}
\label{eqn:57}
T_{i,k}(\nu,Q^2,m^2)&=& \frac{1}{\pi} \int_{Q^2/2}^{\infty} d \nu'
\frac{H_{i,k}(\nu',Q^2,m^2)}{\nu' - \nu}
\nonumber\\
&& = \frac{z}{\pi} \int_0^1 \frac{dz'}{z'} \frac{H_{i,k}(z',Q^2,m^2)}{z-z'}
\nonumber\\
&& = \frac{1}{\pi} \sum_{n=1}^{\infty} z^{-n} \int_0^1 dz' z'^{n-1} 
H_{i,g}(z',Q^2,m^2)\,.
\end{eqnarray}
%------------------
%fig14
%----------------------------------
The graph for $T_{i,g}^{(1)}$ in Fig. \ref{fig:14} 
\footnote{Notice that the lowest order Feynman graphs, shown in this section,
are not complete. Current conservation and gauge invariance require additional
graphs.}
yields the following integral
\begin{eqnarray}
\label{eqn:58}
 T_{i,g}^{(1)}&=& g^2 P_i^{\mu\nu} \epsilon^{\lambda}(p)\epsilon^{\sigma}(p) 
\nonumber\\
&& \times \int \frac{d^N k}{(2\pi)^N} \frac{\gamma_{\lambda}(\ks + m)
\gamma_{\mu} ( \ks + \qs + m) \gamma_{\nu} (\ks + m) \gamma_{\sigma}}
{[(k-p)^2-m^2](k^2-m^2)^2 [(k + q)^2 - m^2]}
\nonumber\\
&\equiv& \int d^N k \, f(k,q,p,m)\,.
\end{eqnarray}
where $P_i^{\mu\nu}$ is a projection operator with $i=2,L$ and 
$\epsilon^{\lambda}(p)$ stands for the polarization vector of the gluon.
In the limit $Q^2 \gg m^2$ the above expression will be called
$T_{i,g}^{{\rm ASYMP},(1)}$ which can be split into the coefficient function
\begin{eqnarray}
\label{eqn:59}
&& \frac{1}{\pi} \sum_{n=1}^{\infty} z^{-n} \int_0^1 dz' z'^{n-1} 
{\cal C}_{i,g}^{(1)}\Big(z',\frac{Q^2}{\mu^2}\Big)=
\nonumber\\
&& \Big [ \mathop{\mbox{lim}}\limits_{\vphantom{\frac{A}{A}} Q^2 \gg m^2} 
\int d^N k \, f(k,q,p,m) - \int d^N k \,
\mathop{\mbox{lim}}\limits_{\vphantom{\frac{A}{A}} Q^2 \gg m^2}
 \, f(k,q,p,m) \Big ]\,,
\end{eqnarray}
and the operator matrix element
\begin{eqnarray}
\label{eqn:60}
 \frac{1}{\pi} \sum_{n=1}^{\infty} z^{-n} \int_0^1 dz' z'^{n-1}
A_{cg}^{(1)}\Big(z',\frac{\mu^2}{m^2}\Big)
= \int d^N k \, \mathop{\mbox{lim}}\limits_{\vphantom{\frac{A}{A}} Q^2 \gg m^2}
 \, f(k,q,p,m)\,.
\end{eqnarray}
The first term in $T_{i,g}^{\rm ASYMP,(1)}$ has no singularity at $m = 0$ and 
it carries
the whole $Q^2$-dependence. The absence of the C-singularity can be checked
by taking $k \propto p$. The dependence on $m$ is transferred to $A_{cg}^{(1)}$
which however is independent of $q^2=-Q^2$ as we will show below. 
Further by interchanging limits and integrations in Eqs. (\ref{eqn:59})
and (\ref{eqn:60}) one
introduces an artificial UV singularity which has to be subtracted before
one gets the finite ${\cal C}_{i,g}^{(1)}$ and $A_{cg}^{(1)}$. To evaluate
the expression in Eq. \ref{eqn:60} we make the Taylor expansion
\begin{eqnarray}
\label{eqn:61}
&&\frac{\gamma_{\mu} (\ks + \qs + m) \gamma_{\nu}}{(k+q)^2-m^2}=
\frac{1}{q^2} \sum_{n=1}^{\infty} \ \gamma_{\mu} \qs \gamma_{\nu}
\Big ( \frac{2k\cdot q}{-q^2} \Big )^{n-1}
\nonumber\\
&&= - \sum_{n=1}^{\infty} \frac{2^{n-1}}{(-q^2)^n}
\gamma_{\mu}\gamma_{\mu_1}\gamma_{\nu} k_{\mu_2} \cdots
k_{\mu_n} q^{\mu_1} \cdots q^{\mu_n}\,,
\end{eqnarray}
so that we get
\begin{eqnarray}
\label{eqn:62}
&& \frac{1}{\pi} \sum_{n=1}^{\infty} z^{-n} \int_0^1 dz' z'^{n-1}
A_{cg}^{(1)}\Big(z',\frac{\mu^2}{m^2}\Big) 
= - g^2 \sum_{n=1}^{\infty} \frac{2^{n-1}}
{(-q^2)^n} q^{\mu_1} \cdots q^{\mu_n} 
\nonumber\\
&& \times P_i^{\mu\nu} \epsilon^{\lambda}(p)
\epsilon^{\sigma}(p)
 \int \frac{d^N k}{(2\pi)^N} \frac{\gamma_{\lambda}(\ks + m)
\gamma_{\mu} \gamma_{\mu_1} \gamma_{\nu} (\ks + m) \gamma_{\sigma}}
{[(k-p)^2-m^2](k^2-m^2)^2 } k_{\mu_2} \cdots k_{\mu_n}\,,
\end{eqnarray}
The vertex $2^{n-1}\gamma_{\mu_1} k_{\mu_2} \cdots k_{\mu_n}$ originates
from the heavy quark operator matrix element (OME) given by
\begin{eqnarray}
\label{eqn:63}
\langle c(k) \mid O_{c,\mu_1 \cdots \mu_n}(0) \mid c(k) \rangle\,.
\end{eqnarray}
The heavy quark operator, which is a gauge invariant object 
(physical operator), is defined by
\begin{eqnarray}
\label{eqn:64}
 O_{c,\mu_1 \cdots \mu_n}(x) = \bar \psi(x) \gamma_{\mu_1} D_{\mu_2} \cdots
D_{\mu_n} \psi(x)\,, 
\end{eqnarray}
with the heavy quark (charm) field $\psi$ and the covariant derivative  
$D_{\mu}=\partial_{\mu} + i g A_{\mu}$.\\
If we now sandwich the above operator between physical gluon states one
obtains
\begin{eqnarray}
\label{eqn:65}
&& \langle g(k) \mid O_{c,\mu_1 \cdots \mu_n}(0) \mid g(k) \rangle=
\nonumber\\
&& g^2 \epsilon^{\lambda}(p) \epsilon^{\sigma}(p)
 \int \frac{d^N k}{(2\pi)^N} \frac{\gamma_{\lambda}(\ks + m)
 \gamma_{\mu_1}(\ks + m) \gamma_{\sigma}}
{[(k-p)^2-m^2](k^2-m^2)^2 } k_{\mu_2} \cdots k_{\mu_n}
\nonumber\\
&& = \hat A_{cg}^{(1),n}(\frac{\mu^2}{m^2})[ p_{\mu_1} \cdots p_{\mu_n} 
-g_{\mu_1\mu_2} p_{\mu_3} \cdots p_{\mu_n} 
\nonumber\\
&& -  g_{\mu_1\mu_2}g_{\mu_3\mu_4} 
 p_{\mu_5} \cdots p_{\mu_n} \cdots ]_{\rm symmetric}\,.
\end{eqnarray}
Notice that the projection which survives in Eq. (\ref{eqn:62}) 
is given by $i=2$ since
$P_2^{\mu\nu}$ contains the metric tensor $g_{\mu\nu}$ only. Projections 
containing the vectors $q^{\mu}$ and $p^{\mu}$ vanish because of current
conservation and the on-shell condition $p^2=0$ respectively. Therefore 
$T_{L,g}^{\rm ASYMP,(1)}$ and also $H_{L,g}^{\rm ASYMP,(1)}$ 
have no C-divergence
for $m=0$ and the only contribution comes from ${\cal C}_{L,g}^{(1)}$ which is
$Q^2$ independent. The equality between Eq. (\ref{eqn:62}) and Eq. 
(\ref{eqn:65}) shows that the former represents the OME $A_{cg}^{(1)}$ as 
announced below Eq. (\ref{eqn:55}).
%---------------------------
%fig15
\begin{figure}
\begin{center}
  \begin{picture}(120,80)
  \Gluon(0,0)(30,30){3}{7}
  \ArrowLine(60,80)(30,30)
  \ArrowLine(90,30)(60,80)
  \ArrowLine(30,30)(90,30)
  \Gluon(90,30)(120,0){3}{7}
  \GCirc(60,80){4}{0.3}
    \Text(25,10)[t]{$\uparrow p$}
    \Text(95,10)[t]{$\downarrow p$}
    \Text(38,60)[t]{$k$}
    \Text(83,60)[t]{$k$}
    \Text(60,23)[t]{$k+p$}
  \end{picture}
\hspace*{1cm}
  \begin{picture}(120,80)
  \Gluon(0,0)(30,30){3}{7}
  \ArrowLine(30,30)(60,80)
  \ArrowLine(60,80)(90,30)
  \ArrowLine(90,30)(30,30)
  \Gluon(90,30)(120,0){3}{7}
  \GCirc(60,80){4}{0.3}
    \Text(25,10)[t]{$\uparrow p$}
    \Text(95,10)[t]{$\downarrow p$}
    \Text(38,60)[t]{$k$}
    \Text(83,60)[t]{$k$}
    \Text(60,23)[t]{$k-p$}
\end{picture}
  \caption{Operator graphs corresponding with the operator matrix
           element $A_{cg}^{(1)}$.}
  \label{fig:15}
\end{center}
\end{figure}
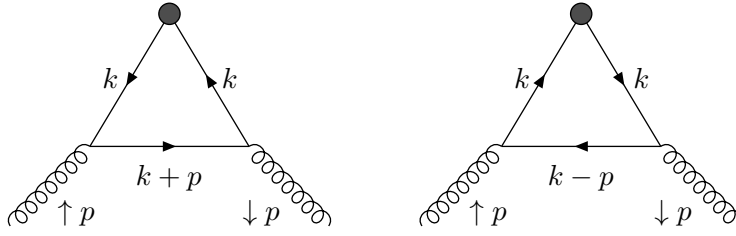
%-----------------------------
It is now easy to extract the OME if Eq. (\ref{eqn:65}) is contracted by 
the tensor $\Delta_{\mu_1}\Delta_{\mu_2} \cdots \Delta_{\mu_n}$ where
$\Delta_{\mu}$ is a lightlike vector i.e. $\Delta^2=0$. Hence we get
the expression 
\begin{eqnarray}
\label{eqn:66}
 \hat A_{cg}^{(1),n}\Big(\frac{\mu^2}{m^2}\Big)
= \int_0^1 d z' z'^{n-1} a_s \Big [
P_{qg}^{(0)}(z') \Big ( \frac{1}{\epsilon_{UV}} + \frac{1}{2} 
\ln(\frac{\mu^2}{m^2})
\Big ) + a_{qg}^{(1)}(z') \Big ]\,,
\end{eqnarray}
with
\begin{eqnarray}
\label{eqn:67}
 A_{cg}^{(1)}\Big(z,\frac{\mu^2}{m^2}\Big) = \frac{1}{\pi} \sum_{n=1}^{\infty}
z^{-n} A_{cg}^{(1),n}\Big(\frac{\mu^2}{m^2}\Big)\,.
\end{eqnarray}
After subtraction of the UV pole term, for which we choose the 
$\overline{\rm MS}$-scheme, one obtains the finite OME in Eq. (\ref{eqn:54}).
The finite
coefficient function ${\cal C}_{2,g}^{(1)}$ Eq. (\ref{eqn:53}) is obtained from
Eq. (\ref{eqn:59}) in a similar way.\\
Generalizing mass factorization to higher orders one can write
\begin{eqnarray}
\label{eqn:68}
H_{i,k}^{\rm ASYMP}(Q^2,m^2)=A_{lk}\Big(\frac{\mu^2}{m^2}\Big) 
\otimes {\cal C}_{i,l}
\Big(\frac{Q^2}{\mu^2} \Big)\,.
\end{eqnarray}
If we expand the OME's $A_{kl}$ and the light parton coefficient functions
${\cal C}_{i,k}$ up to order $\alpha_s^2$ one obtains the following equations
\begin{eqnarray}
\label{eqn:69}
H_{i,g}^{\rm ASYMP,(1)}=A_{cg}^{(1)}\otimes {\cal C}_{i,c}^{(0)} + 
A_{gg}^{(0)}\otimes {\cal C}_{i,g}^{(1)}\,,
\end{eqnarray}
\begin{eqnarray}
\label{eqn:70}
H_{i,g}^{\rm ASYMP,(2)}=A_{cg}^{(2)}\otimes {\cal C}_{i,c}^{(0)} + 
A_{cg}^{(1)} \otimes {\cal C}_{i,c}^{(1)}+
A_{gg}^{(0)}\otimes {\cal C}_{i,g}^{(2)}\,,
\end{eqnarray}
\begin{eqnarray}
\label{eqn:71}
H_{i,q}^{\rm ASYMP,(2)}=A_{cq}^{(2)}\otimes {\cal C}_{i,c}^{(0)} + 
A_{qq}^{(0)}\otimes {\cal C}_{i,q}^{(2)}\,.
\end{eqnarray}
Here $A_{gg}$ is obtained by sandwiching the gauge invariant gluonic operator
\begin{eqnarray}
\label{eqn:72}
O_{g,\mu_1 \cdots \mu_n} (x) 
= F_{\mu_1}^{\alpha} D_{\mu_2} \cdots D_{\mu_{n-1}}
F_{\mu_n,\alpha}\,,
\end{eqnarray}
between physical gluon states
The heavy quark coefficient functions $L_{i,k}$ have to be added to 
${\cal C}_{i,k}$ before we can apply mass factorization
\begin{eqnarray}
\label{eqn:73}
{\cal C}_{i,k}\Big(n_f,\frac{Q^2}{\mu^2}\Big) + L_{i,k}^{\rm ASYMP}(Q^2,m^2) =
A_{lk,c}\Big(\frac{\mu^2}{m^2}\Big) \otimes 
{\cal C}_{i,l}\Big(n_f+1,\frac{Q^2}{\mu^2}\Big)\,.
\end{eqnarray}
This relation also involves a redefinition of the strong coupling constant
which in the right-hand-side now depends on $n_f+1$ instead of $n_f$.
The main effect of this operation is that the number of light flavours $n_f$
appearing in ${\cal C}_{i,k}$ is enhanced by one unit.
The $A_{kl,c}$ are given by all light parton OME's containing charm loop 
contributions to the gluon self energy only. Up to order $\alpha_s^2$ we have
\begin{eqnarray}
\label{eqn:74}
L_{i,k}^{\rm ASYMP,(2)}=A_{qq,c}^{(2)} \otimes {\cal C}_{i,q}^{(0)} + 
A_{qq}^{(0)} \otimes {\cal C}_{i,q}^{(2)} + a_s \beta_{0,c}\, 
{\cal C}_{i,q}^{(1)} \ln(\frac{m^2}{\mu^2})\,.
\end{eqnarray}
Finally we want to emphasize that beyond the second order the above
mass factorization relations become much more complicated. Here the heavy
quark OME's $A_{ck}$ can also contribute to Eq. (\ref{eqn:73})
whereas the OME's $A_{kl,c}$ also enter Eq. (\ref{eqn:68}).
Using the mass factorization relations in Eqs. (\ref{eqn:69})-(\ref{eqn:71})
we have computed the asymptotic heavy quark coefficient
functions $H_{i,k}^{\rm ASYMP,(2)}$ for $k=q,g$ and $i=2,L$ in \cite{bmsmn}. 
The same was done for $L_{i,q}^{\rm ASYMP,(2)}$ Eq. (\ref{eqn:74}). 
This was possible because
the light parton coefficient functions and the heavy quark OME's were already
known in \cite{zn}  and \cite{bmsmn} respectively. 
The expressions reveal the following characteristic behaviour
\begin{eqnarray}
\label{eqn:75}
H_{i,k}^{{\rm ASYMP},(l)}(z,Q^2,m^2,\mu^2)\sim \alpha_s^l 
\sum_{n+j\leq l} a_{nj}(z)
\ln^n\Big(\frac{\mu^2}{m^2}\Big)\ln^j\Big(\frac{Q^2}{m^2}\Big)\,,
\end{eqnarray}
with an analogous expression for $L_{i,k}^{\rm ASYMP}$. 
One of the most important
features, discovered in \cite{bmsmn}, is that the asymptotic 
expressions tend to
the exact heavy quark coefficient functions at rather low $\xi$-values 
(see Eq. (\ref{eqn:43}))
provided $z$ is not too small. According to ref. \cite{bmsmn} 
$\xi=10$ for $z < 0.01$
which in charm production corresponds to $Q^2=22.5 \,({\rm GeV/c})^2$ with 
$m=1.5~{\rm GeV/c}$ The consequences of this behaviour for the charm component 
of the structure function will be discussed in the next section.

\section{Description of $F_{i,c}$ in the various schemes}
In the literature one has proposed various descriptions of the charm component
of the structure function in the framework of extrinsic charm production.
Here one can distinguish the following schemes.
\begin{itemize}
\item[1] \underline{The three flavour number scheme (TFNS)}\\[3mm] 
Here the production mechanisms are given by the photon-gluon fusion process  
and the higher order reactions. The charm component of the structure  
function $F_{i,c}$ is given by Eq. (\ref{eqn:37}). The parton densities are  
given by the three light flavour densities $u,d,s$ and the gluon density 
$g$.
\item[2] \underline{The four flavour number scheme (FFNS)}\\[3mm]
In this case $F_{i,c}$ is expressed into convolutions of light parton
coefficient functions with light parton densities. The latter are represented
by four light flavours, which includes the charm quark, and the gluon.
\item[3] \underline{The variable-flavour number scheme (VFNS)}\\[3mm]
This scheme interpolates between the results of the structure functions
$F_{i,c}$ obtained from the TFNS and the FFNS.
\end{itemize}
%---------------------
%fig16
\begin{figure}
\begin{center}
  \begin{picture}(120,120)
    \DashArrowLine(0,20)(25,20){5}
    \DashArrowLine(105,20)(130,20){5}
    \DashArrowLine(25,20)(65,90){5}
    \DashArrowLine(65,90)(105,20){5}
   \ArrowArc(65,20)(15,0,180)
   \ArrowArc(65,20)(15,180,360)
  \Gluon(25,20)(50,20){3}{4}
  \Gluon(80,20)(105,20){3}{4}
  \Photon(65,120)(65,90){3}{4}
  \end{picture}
  \caption{Two-loop vertex correction containing a heavy quark loop.
          It contributes to ${\cal C}_{i,q}$.}
  \label{fig:16}
\end{center}
\end{figure}
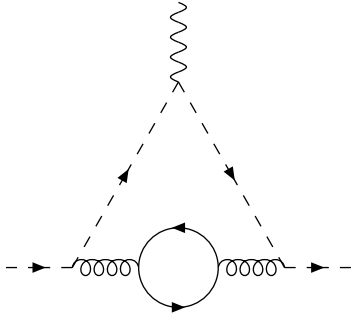
%-------------------------------
In this section we will now derive the expressions for $F_{i,c}$ in the FFNS
and the VFNS from the TFNS .\\
In order to derive $F_{i,c}^{\rm FFNS}$ we need the deep inelastic structure 
function which receives
contributions from subprocesses containing light partons in the initial and 
final state only. It is given by
\begin{eqnarray}
\label{eqn:76}
&& F_i(n_f,Q^2) = 
\nonumber\\
&& \frac{1}{n_f} \sum_{k=1}^{n_f} e_k^2 \Big [ f_q^{\rm S}(n_f,\mu^2)
 \otimes {\cal C}_{i,q}^{\rm S}\Big(n_f,\frac{Q^2}{\mu^2}\Big)
+ f_g^{\rm S}(n_f,\mu^2) \otimes {\cal C}_{i,g}^{\rm S}
\Big(n_f,\frac{Q^2}{\mu^2}\Big) 
\nonumber\\
&& + n_f f_k^{\rm NS}(n_f,\mu^2) \otimes 
{\cal C}_{i,q}^{\rm NS}\Big(n_f,\frac{Q^2}{\mu^2}\Big)
\Big ]\,.
\end{eqnarray}
In the case of TFNS we have to choose $n_f=3$.
In the above expression the coefficient functions ${\cal C}_{i,k}$ contain
heavy quark loop contributions to the gluon self energies appearing in the 
light partonic matrix elements. This happens for the first time in order
$\alpha_s^2$. An example is shown in Fig. \ref{fig:16}. The charm quark loop
contributions will now be removed from $F_i(n_f,Q^2)$ and added to $F_{i,c}$ 
in (\ref{eqn:37}). In the subsequent part of this section the latter
will be called $F_{i,c}^{\rm EXACT}$. In this way the coefficient functions 
$L_{i,k}$ get the proper asymptotic behaviour when
$Q^2 \gg m^2$ as indicated in Eq. (\ref{eqn:75}). This allows us to perform 
mass factorization after we have added the light parton coefficient function
${\cal C}_{i,k}$ in which all contributions from heavy quark loops are removed 
(see Eq. (\ref{eqn:73})). Next we define the following quantity
\begin{eqnarray}
\label{eqn:77}
F_{i,c}^{\rm ASYMP}(n_f,x,Q^2, m^2 )= \lim_{Q^2 \gg m^2}
\Big[F_{i,c}^{\rm EXACT}(n_f,x,Q^2,m^2)\Big] \,.
\end{eqnarray}
%----------------------------------
%fig17
\begin{figure}
 \begin{center}
  {\unitlength1cm
   \epsfig{file=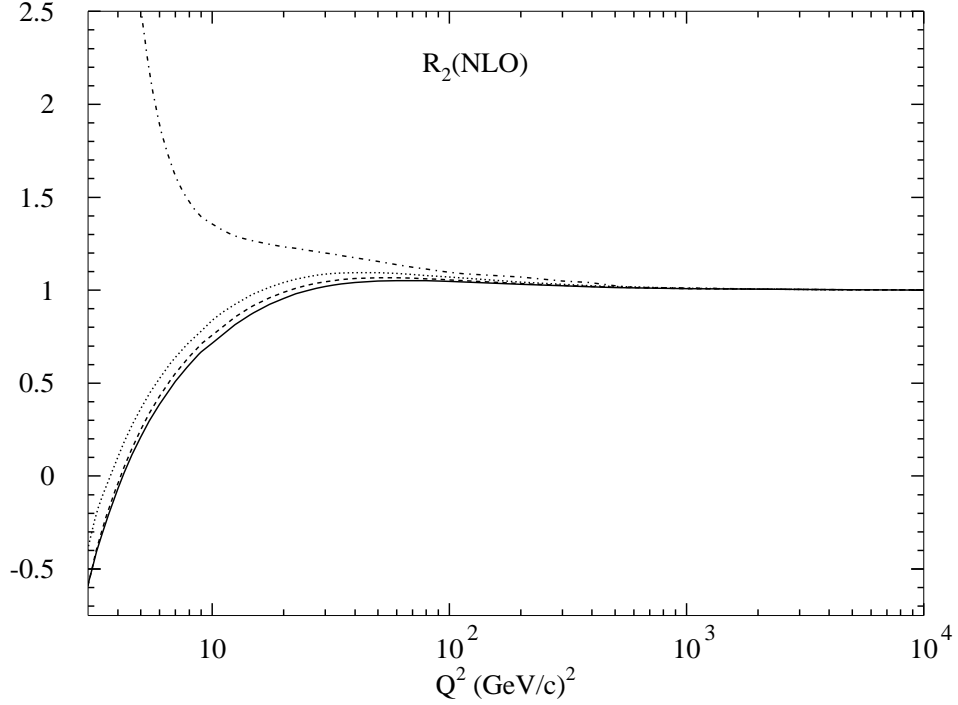,bbllx=520pt,bblly=95pt,bburx=105pt,bbury=710pt,%
      height=8cm,width=12cm,clip=,angle=270}
    }  
 \end{center}
\caption{$R_2({\rm NLO})$ plotted as a function of $Q^2$ at fixed $x$; 
$x=0.1$ (dashed-dotted line), $x=0.01$ (dotted line), $x=10^{-3}$ (dashed line)
, $x=10^{-4}$ (solid line).}
\label{fig:17}
\end{figure}
%-----------------------------------------
Notice that $F_{i,c}^{\rm ASYMP}$ is given by the same expression as
$F_{i,c}^{\rm EXACT}$ except that now the exact heavy quark coefficient
functions are replaced by their asymptotic analogues which have the form 
presented in Eq. (\ref{eqn:75}). In Fig. \ref{fig:17} we have plotted in  
next-to-leading order (NLO) the ratio
\begin{eqnarray}
\label{eqn:78}
 R_2(x,Q^2,m_c^2)=
\frac{F_{2,c}^{\rm ASYMP}(x,Q^2,m_c^2)}{F_{2,c}^{\rm EXACT}(x,Q^2,m_c^2)}
\,.
\end{eqnarray}
For this plot we have adopted the parton density set 
with $\Lambda_4= 200$ MeV from 
\cite{grv94} ($\overline{\rm MS}$-scheme). The mass factorization scale 
is chosen to be $\mu=Q$. The charm quark mass is equal to $m_c=1.5~{\rm GeV/c}$.
{}From this figure we infer that for $Q^2 > 20~({\rm GeV/c})^2$ 
and $x<0.01$, $F_{2,c}^{\rm ASYMP}$ coincides with $F_{2,c}^{\rm EXACT}$.
For $F_{L,c}$ this happens when $Q^2$ is much larger i.e.  
$Q^2 > 1000~({\rm GeV/c})^2$.
This implies that the large logarithmic terms in  
the heavy quark coefficient functions $H_{i,k}$ and $L_{i,k}$ as given by
Eq. (\ref{eqn:75}) entirely determine the charm component of the 
structure function. 
Since these corrections vitiate the
perturbation series when $Q^2$ gets large they should be resummed in all 
orders of perturbation theory. This procedure has been carried out in
\cite{bmsn} and it consists of four steps. First we add $F_i(n_f,Q^2)$ in 
Eq. (\ref{eqn:76}) to $F_{i,c}^{\rm ASYMP}(n_f,x,Q^2,m^2)$ in 
Eq. (\ref{eqn:77}) and choose $n_f=3$.
Second we apply mass 
factorization to the asymptotic heavy quark coefficient functions according
to Eqs. (\ref{eqn:68}),(\ref{eqn:73}). 
%----------------------------
%fig18
\begin{figure}
 \begin{center}
  {\unitlength1cm
   \epsfig{file=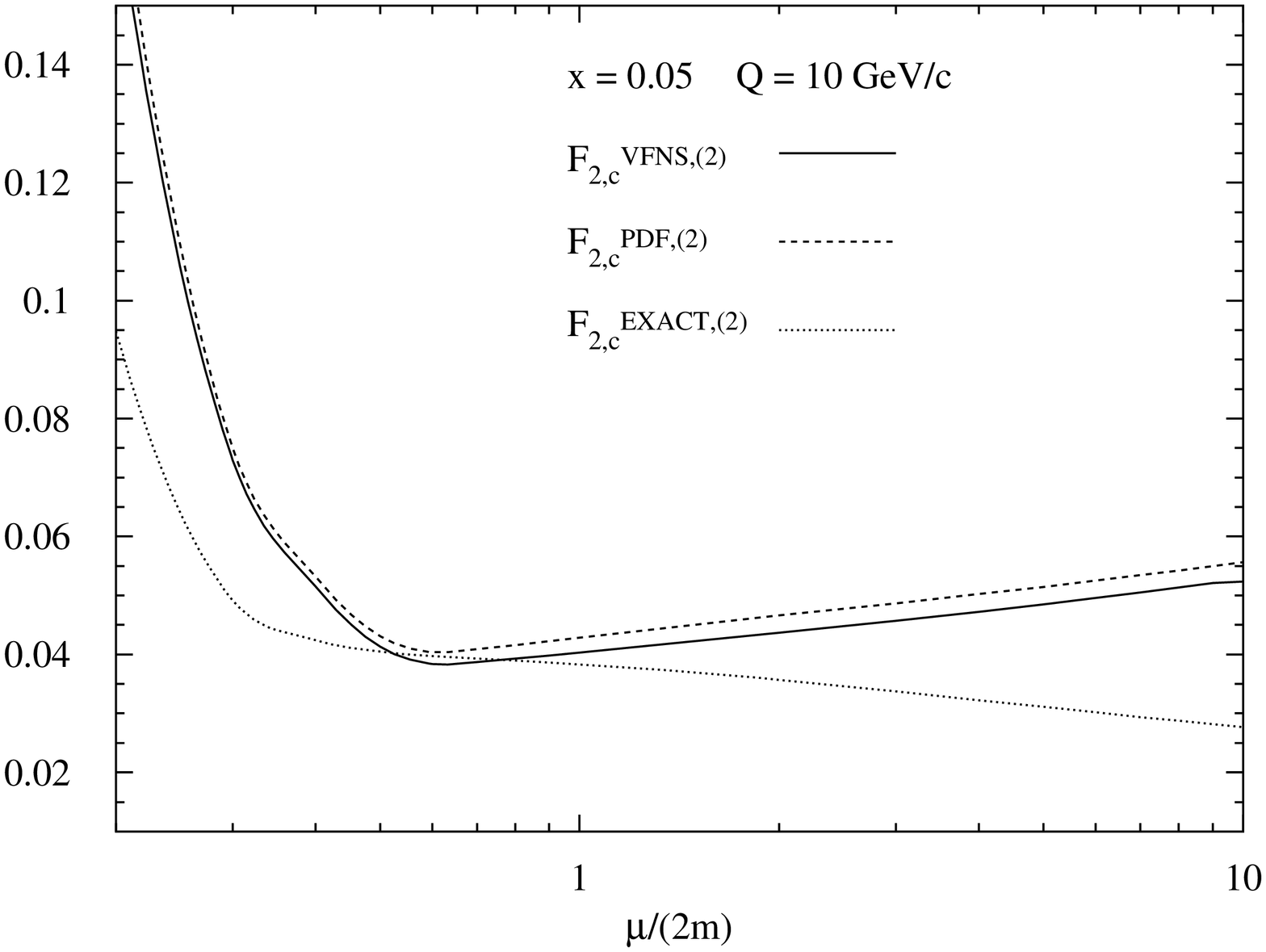,bbllx=520pt,bblly=95pt,bburx=105pt,bbury=710pt,%
      height=8cm,width=12cm,clip=,angle=270}
     }
   \end{center}
\caption{Scale dependence of the structure functions at $x=0.05$ and
$Q=10~{\rm GeV/c}$; $F_{2,c}^{\rm VFNS,(2)}$ (solid line), 
$F_{2,c}^{\rm PDF,(2)}$ (dashed line), $F_{2,c}^{\rm EXACT,(2)}$ 
(dotted line).}
\label{fig:18}
\end{figure}
%-------------------------------------------------------
In the third step we define new parton densities which are now
presented in a four flavour number scheme (FFNS). 
For the three light flavour densities $k=u,d,s$ we have
\begin{eqnarray}
\label{eqn:79}
f_k(4, \mu^2) + f_{\bar k}(4, \mu^2)\!\! &=& \!\!
 A_{qq,c}^{\rm NS}\Big(3, \frac{\mu^2}{m^2}\Big)\otimes
\Big[ f_k(3, \mu^2) + f_{\bar k}(3, \mu^2) \Big]
\nonumber \\  &&
+ \tilde A_{qq,c}^{\rm PS}\Big(3, \frac{\mu^2}{m^2}\Big) \otimes
\Sigma(3, \mu^2)
\nonumber \\ &&
+ \tilde A_{qg,c}^{\rm S}\Big(3, \frac{\mu^2}{m^2}\Big) \otimes
G(3, \mu^2)\,.
\end{eqnarray}
The gluon density in the FFNS reads 
\begin{eqnarray}
\label{eqn:80}
 G(4, \mu^2) =  A_{gq,c}^{\rm S}(3,\mu^2)
\otimes \Sigma(3,\mu^2)
+ A_{gg,c}^{\rm S}(3,\mu^2)\otimes G(3,\mu^2)\,.
\end{eqnarray}
Finally we get a new light flavour density represented by the charm quark
\begin{eqnarray}
\label{eqn:81}
f_{c+\bar c}(4, \mu^2) &\equiv&  f_{4}(4, \mu^2)
+ f_{\overline {4}}(4, \mu^2)
\nonumber \\
&=&  A_{cq}^{\rm S}\Big(3, \frac{\mu^2}{m^2}\Big)\otimes
\Sigma(3, \mu^2)
+  A_{cg}^{\rm S}\Big(3, \frac{\mu^2}{m^2}\Big) \otimes
G(3, \mu^2)\,.
%\nonumber\\
\end{eqnarray}
The above charm quark density has the property that it
does not vanish at $\mu = m$ in the $\overline{\rm MS}$-scheme contrary to
what is usually assumed in the literature (see e.g. \cite{grv92},
\cite{cteq}, \cite{mrs}), \cite{lt}). In the fourth step we rearrange
terms and obtain
\begin{eqnarray}
\label{eqn:82}
F_{i,c}^{\rm ASYMP}(n_f,x,Q^2,m^2)+F_i(n_f,x,Q^2)=F_i(n_f+1,x,Q^2)\,,
\end{eqnarray}
which is the FFNS result for the total structure function. From the latter 
quantity one can extract the expression for the charm quark component of
the proton structure function in the FFNS which will be denoted by
\begin{eqnarray}
\label{eqn:83}
&& F_{i,c}^{\rm PDF} (n_f+1,Q^2)   =  
\nonumber\\
&& e_c^2 \Big[f_{c +\bar c}(n_f+1,\mu^2) 
\otimes {\cal C}_{i,q}^{\rm NS}\Big(n_f+1,\frac{Q^2}{\mu^2}\Big)
+ \Sigma(n_f+1,\mu^2) \otimes
\nonumber\\
&& \tilde{\cal C}_{i,q}^{\rm PS}\Big(n_f+1,\frac{Q^2}{\mu^2}\Big)
+ G(n_f+1,\mu^2) \otimes \tilde{\cal C}_{i,g}^{\rm S}
\Big(n_f+1,\frac{Q^2}{\mu^2}\Big) \Big]\,,
\end{eqnarray}
where we have defined
\begin{eqnarray}
\label{eqn:84}
{\cal C}_{i,q}^{\rm S}(n_f)={\cal C}_{i,q}^{\rm NS}(n_f)
+n_f \tilde{\cal C}_{i,q}^{\rm PS}(n_f) \,, \qquad
{\cal C}_{i,g}^{\rm S}=n_f\tilde{\cal C}_{i,g}^{\rm S} \,.
\end{eqnarray} 
The superscript PDF in Eq. (3) stands for parton density function which means
that the charm component of the structure function is completely
expressed into parton densities multiplied by the light parton coefficient
functions. Notice that $F_{i,c}^{\rm PDF}$ is a renormalization group
invariant like $F_{i,c}^{\rm EXACT}$ and $F_{i,c}^{\rm ASYMP}$ so that
they satisfy the equation $D~F_{i,c}=0$ (see Eq. (\ref{eqn:38}) and below).
Further $F_{i,c}^{\rm PDF}$ originates from the charm quark coefficient 
functions $H_{i,k}$ from which it follows that
the former is proportional to $e_c^2$ only. 
The functions $L_{i,k}$, which are
multiplied by the light charge squared $e_k^2$, contribute to both
$F_{i,c}^{\rm PDF}(n_f+1,Q^2)$ and 
$F_2(n_f+1,Q^2)$ so that the number of flavours in these structure functions
are increased by one unit.
The FFNS charm quark density is mainly determined by
the size of the TFNS gluon density $G(n_f,z,\mu^2)$ for $n_f=3$. 
Therefore the latter plays a major role in
the behaviour of $F_{i,c}^{\rm EXACT}$  
as well as of $F_{i,c}^{\rm PDF}$ . 
An analysis of both structure functions in \cite{bmsn} reveals
that the former gives the best description of charm production in the
threshold region where $Q^2$ is small and $x$ is large. On the other hand
when $Q^2$ is large and $x$ is small it turns out that it is better to use
$F_{2,c}^{\rm PDF}$ because it is in this region where the large logarithms
in Eq. (\ref{eqn:75}) dominate so that they have to be resummed.
Therefore the TFNS is the most suitable scheme for the charm
component of the structure function near threshold whereas far away from this
region it turns out that the FFNS is more appropriate. 

%\newpage
%--------------------------
%fig19
\begin{figure}
 \begin{center}
 {\unitlength1cm
  \epsfig{file=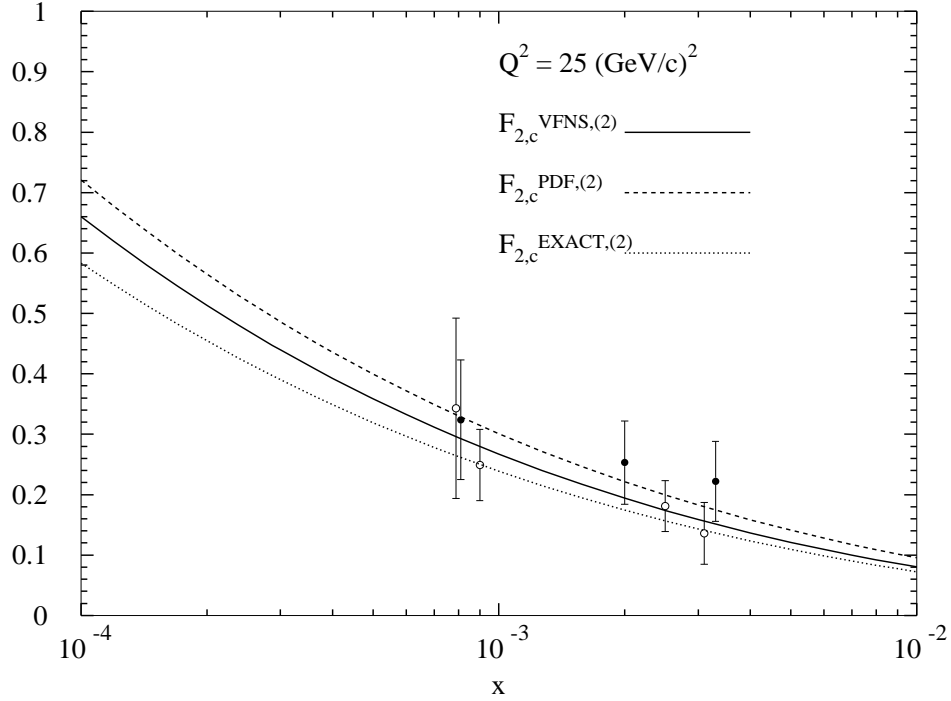,bbllx=520pt,bblly=95pt,bburx=105pt,bbury=710pt,%
      height=8cm,width=12cm,clip=,angle=270}
   }
  \end{center}
\caption{Structure functions at $Q^2=25~({\rm GeV/c})^2$;
$F_{2,c}^{\rm VFNS,(2)}$ (solid line), $F_{2,c}^{\rm PDF,(2)}$ (dashed line),
$F_{2,c}^{\rm EXACT,(2)}$ (dotted line). The experimental data are from
[6] (closed circles) and [7] (open circles).}
\label{fig:19}
\end{figure} 
%---------------------------------------
One also needs a scheme which merges the advantages of these 
two pictures and provides us with good description of $F_{2,c}$ 
in the intermediate regime in $Q^2$. This is
given by the so called variable flavour number scheme (VFNS). 
In \cite{bmsn} we proposed the following VFNS structure function
\begin{eqnarray}
\label{eqn:85}
F_{i,c}^{\rm VFNS}(x,Q^2,m^2) &=&
F_{i,c}^{\rm PDF}(n_f+1,x,Q^2)
+ F_{i,c}^{\rm EXACT}(n_f,x,Q^2,m^2) \nonumber \\ &&
- F_{i,c}^{\rm ASYMP}(n_f,x,Q^2,m^2)\,.
\end{eqnarray}
The above expression is a generalization of Eq. (9) in \cite{acot}, which
was only presented in leading order (LO) and has been implemented in a recent
global parton density analysis \cite{lt}. 
(A different VFNS scheme has recently been proposed in
\cite{mrrs}.)  
In LO the VFNS scheme has the properties
that for $Q^2 \gg m^2$, $F_{2,c}^{\rm EXACT} \rightarrow F_{2,c}^{\rm ASYMP}$
which means that $F_{2,c}^{\rm VFNS} \rightarrow F_{2,c}^{\rm PDF}$.
Further at low
$Q^2$ (i.e. $Q^2 \leq m^2$) $F_{2,c}^{\rm ASYMP} \rightarrow 
F_{2,c}^{\rm PDF}$ so that $F_{2,c}^{\rm VFNS} \rightarrow 
F_{2,c}^{\rm EXACT}$.
provided we put $z_{\rm max}=1$ in $F_{2,c}^{\rm ASYMP}$.
However the last relation is no longer true in higher order in $\alpha_s$.
The main reason is that new production mechanisms appear in NLO giving rise 
to the coefficient functions $L_{i,k}$. The latter contain the 
prefactor $e_k^2$ (e.g. the Compton process) and show up
in $F_{2,c}^{\rm EXACT}$ and $F_{2,c}^{\rm ASYMP}$ but not in
$F_{2,c}^{\rm PDF}$, which is proportional to $e_c^2$ only (see Eq. 
(\ref{eqn:83}). This higher order effect is only noticeable in the threshold 
regime where $x$ is very large and $Q^2$ is very small.
%--------------------------
%fig20
\begin{figure}
 \begin{center}
 {\unitlength1cm
 \epsfig{file=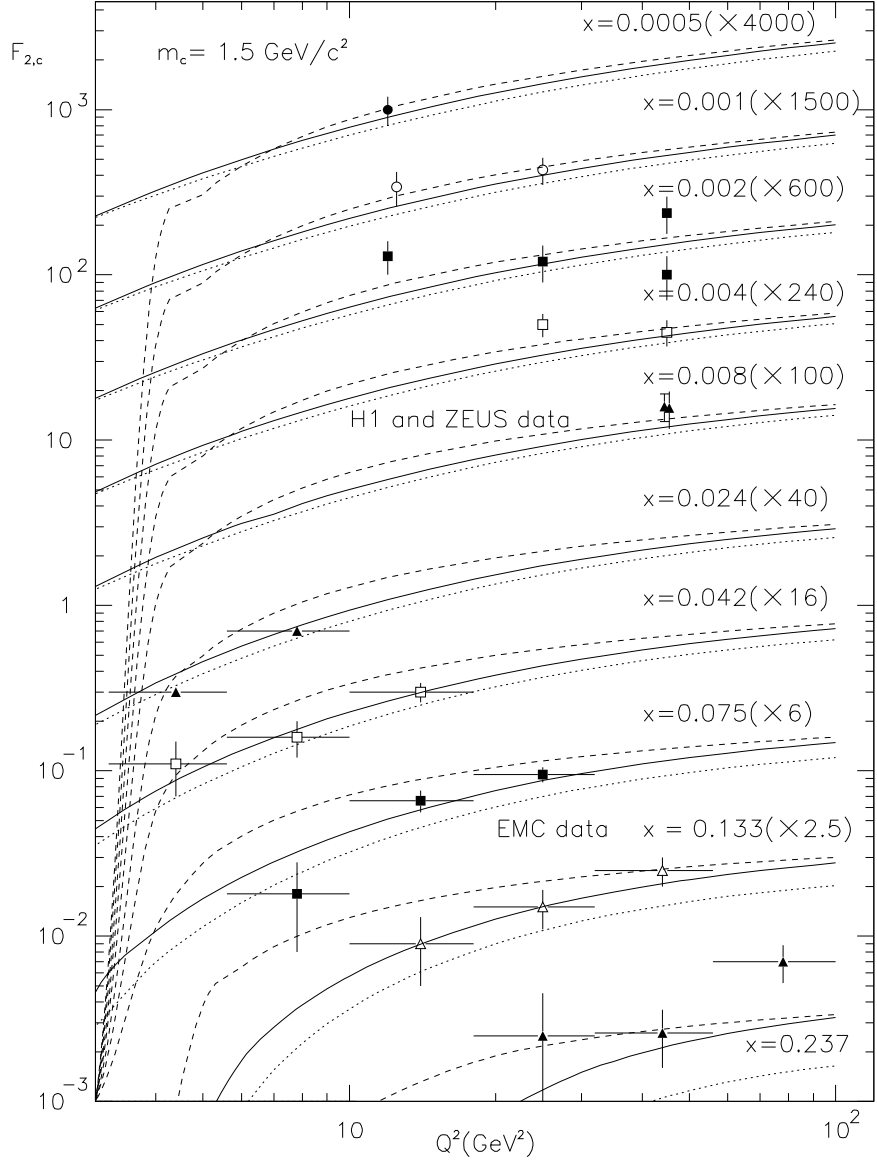,bbllx=50pt,bblly=100pt,bburx=510pt,bbury=720pt,height=16cm,width=12cm}
   }
  \end{center}
\caption{Structure functions
$F_{2,c}^{\rm VFNS,(2)}$ (solid line), $F_{2,c}^{\rm PDF,(2)}$ (dashed line),
$F_{2,c}^{\rm EXACT,(2)}$ (dotted line). The experimental data are from
[27], [6] and [7].}
\label{fig:20}
\end{figure} 
%---------------------------------------

As an application we have studied the charm component of the proton structure
function in the three schemes mentioned above. The coefficient functions used
for these structure functions are all computed up to order $\alpha_s^2$ 
(see \cite{lrsn1}, \cite{bmsmn}, \cite{zn}) so that we will denote them by
$F_{2,c}^{\rm EXACT,(2)}$ (Eq. (\ref{eqn:37})), $F_{2,c}^{\rm PDF,(2)}$ 
(Eq. (\ref{eqn:83})) and $F_{2,c}^{\rm VFNS,(2)}$ (Eq. (\ref{eqn:85})).
In Fig. \ref{fig:18} we have plotted the scale ($\mu$)
dependence up to NLO of the structure functions in the different schemes 
mentioned above. For these plots we used the parton densities presented
in the $\overline{\rm MS}$-scheme with 
$\Lambda_4=200$ MeV in \cite{cteq}.
The scale $\mu$ is adopted from \cite{acot} and it is given by
\begin{eqnarray}
\label{eqn:86}
\mu^2 \!\!\!\!\!\!&&
=\,\,\,m^2+ k Q^2 (1- m^2/Q^2)^n \quad \mbox{for} \quad Q^2 > m^2\,,
\nonumber\\
&& =\,\,\, m^2 \quad \mbox{for} \quad Q^2 \leq m^2\,,
\end{eqnarray}
with $k=0.5$, $n=2$ and $m = 1.5~({\rm GeV}/c^2)$.
In \cite{acot} and in \cite{or} it was shown that $F_{2,c}^{\rm VFNS}$
in LO is less sensitive to variations in the scale $\mu$ than each
term on the right-hand-side of Eq. (\ref{eqn:85}) separately. However in NLO
we observe in Fig. \ref{fig:18} that there is no reason to prefer one
scheme over the other. This is corroborated by the findings in \cite{grs} and 
\cite{vogt} where one could show that there is a considerable improvement
in $F_{2,c}^{\rm EXACT,(2)}$ with respect to variations in the mass 
factorization scale when this quantity is computed up to NLO.
In Fig. \ref{fig:19} we plot $F_{2,c}$ in the three different schemes
and compare the results with the recent data
from the H1 \cite{h1} and ZEUS \cite{zeus} collaborations.
Further we have chosen the parton density set in \cite{grv92} 
with $\Lambda_4= 200$ MeV. Notice that all parton density sets in \cite{lt},
\cite{cteq}, \cite{grv92} used for the
plots of the structure functions are presented in FFNS although in
principle one has to choose a TFNS
parametrization for both the parton densities (see e.g
\cite{grv94}) and the running coupling constant for the computation of
$F_{2,c}^{\rm EXACT}$ and $F_{2,c}^{\rm ASYMP}$. However we have checked that
the latter are not significantly altered when 
we replace the parton densities in \cite{grv94}
by those in \cite{grv92}. From Fig. \ref{fig:19} we infer that the data
are in agreement with all schemes in which the structure functions are 
computed. Further the results for $F_{2,c}^{\rm VFNS,(2)}$ are always between
$F_{2,c}^{\rm EXACT,(2)}$ and $F_{2,c}^{\rm PDF,(2)}$. 
%\newpage
It is clear that one
needs more precise data in finer bins of $x$ and $Q^2$ to discriminate 
between the various schemes. Finally we have made a comparison with the data
in the small $x$-region as well as in the large $x$-region obtained by the
experimental groups H1 \cite{h1}, ZEUS \cite{zeus} and EMC \cite{emc} 
respectively. The plots for the various schemes are shown in Fig. \ref{fig:20}.
The agreement between the data and the predictions from the structure
functions in Eq. (\ref{eqn:85}) 
is fairly good except for $x=0.237$. The discrepancy occuring in the threshold
region is mainly due to the large negative contribution coming from Fig. 
\ref{fig:16} and the chosen scale in Eq. (\ref{eqn:86}) which originates from
\cite{acot}. It turns out that at low $Q^2$ this scale becomes too small so
that the running coupling constant is too big. In this region all schemes
lead to negative structure functions in particular for $F_{2,c}^{\rm EXACT}$.
Therefore perturbation theory breaks down and one has to choose a larger
scale. This phenomenon, which was not observed in the LO analysis in 
\cite{acot}, only appears when the NLO corrections are taken into
account.\\[5mm] 
\noindent
Acknowledgements\\

We would like to thank J. Smith and Y. Matiounine for the careful reading of
the manuscript and for giving us some useful comments.

\end{document}